\documentclass[onecolumn,showpacs,amsmath,amssymb]{revtex4}
\usepackage{graphicx}
\usepackage{bm}
\begin{document}
\title{Entanglement Dynamics and Spin Squeezing of The non-linear Tavis-Cummings model
mediated by a Nonlinear Binomial Field}
\author{M. S. Ateto\footnote{E-mail: omersog@yahoo.com, mohamed.ali11@sci.svu.edu.eg}}
\affiliation{Mathematics Department, Faculty of Science at Qena, South Valley
University, 83523 Qena, Egypt
}
%\date{\today}
\begin{abstract}
We show that spin squeezing implies entanglement for quantum tripartite-state, where the subsystem of the bipartite-state is identical. We study the relation between spin squeezing parameters and entanglement through the quantum entropy of a system starts initially in a pure state when the cavity is binomial. We show that spin squeezing can be a convenient tool to give some insight into the subsystems entanglement dynamics when the bipartite subsystem interacts simultaneously with the cavity field subsystem, specially when the interaction occurs off-resonantly without and with a nonlinear medium contained in the cavity field subsystem. We illustrate that, in case of large off-resonance interaction, spin squeezing clarifies the properties of entanglement almost with full success. However, it is not a general rule when the
cavity is assumed to be filled with a non-linear medium. In this case, we illustrate that the insight into entanglement dynamics becomes more clearly in  case of a weak nonlinear medium than in strong nonlinear medium. In parallel, the role of the phase space distribution in quantifying entanglement is also studied. The numerical results of Husimi $Q$-function show that the integer strength of the nonlinear medium produces Schr\"{o}dinger cat states which is necessary for quantum entanglement.
\end{abstract}
\pacs{
03.67.-a,     % Quantum information
03.67.Bg,      % Entanglement production and manipulation
05.30.-d     % Quantum statistical mechanics
 \\
\\
\textit{Accepted in Int. J. Quant. Inf.}}

\maketitle
\section{Overview}
\par The squeezing of the quantum fluctuations is one of the most fundamental
manifestations of the Heisenberg uncertainty relation, which is among the most
important principles of quantum mechanics. 
A long time ago, great effort has been paid to squeezing of the radiation field
due to its strong application in the optical communication~\cite{YS78} and weak
signal detections~\cite{CMC81}. Accordingly, it was established the relationship
between the squeezing of the atoms and that of the radiation field~\cite{WX03}
and the possibility of squeezed atoms to radiate a squeezed field~\cite{POUMO01}.
Moreover, much attention has been devoted to atomic spin squeezing~\cite{MB02,
GB03, SM01, VPG00, WX01, Wx04, YWS07, ZPP91}. Spin squeezed states~\cite{WX01,
VPG00, AALU02, WBI94, KIUE93, AGPU90, LUYEFL02, KUMOPO97, WEMO02, HEYO01, MUZHYO02, POMO01, THMAWI02, GAROBU02, STGEDOMA03, ZHSOLI02, WASOMO01} are quantum correlated states with reduced fluctuations in
one of the collective components. Spin squeezed states offer an interesting possibility for
reducing quantum noise in precision measurements~\cite{WBI94,WBI92,UOMK01,
AALU02, DPJN03, JKP01, MRKSIMW01}, with potentially possible applications in
ultra-precise spectroscopy, atomic interferometers and high precision atomic
clocks~\cite{WBI94,WBI92}.
\par Interestingly, it was found that spin squeezing is closely related to a key
ingredient in quantum information theory and implies quantum
entanglement~\cite{SDCZ01, SORENSEN02, BERSAN02, RAFUSAGU0137}. As there are
various kinds of entanglement, a question arises: what kind of entanglement does
spin squeezing imply? In Ref.~\cite{UOMK01}, it has been found that, for a two-qubit symmetric state, spin squeezing is equivalent to its bipartite entanglement, i.e., spin squeezing implies bipartite entanglement and vice versa. Wang and Sanders~\cite{WASA03}, presented a generalization of the results of Ref.~\cite{UOMK01} to the multiqubit case, where the authors showed that spin squeezing implies pairwise entanglement for arbitrary symmetric multiqubit states. More quantitatively, for spin states with a certain parity, if a state is spin squeezed according to the definition for Kitagawa and Ueda~\cite{KIUE93}, a quantitative relation is obtained between the spin squeezing parameter and the concurrence~\cite{HW97,HW98,XHCY02,ZL03}, quantifying the entanglement of two half-spin particles~\cite{UOMK01,WASA03}.
\par The close relation between the entanglement and spin squeezing enhances the
importance of spin squeezing which motivate us to explore the role of Kerr
medium~\cite{MSAT07,MSAT109,MSAT209,MSAT309} and the nonlinear binomial state on the spin squeezing and
entanglement.
\par A binomial state is one of the important nonclassical states of light which has
attracted much attention in the field of quantum optics in the last few decades,
see for example ~\cite{Buzek90, STSATE85,Vidiella94}. This, in fact, is due to the importance of these states which
have been experimentally produced and characterized. This state is regarded as
one of the intermediate states of the coherent state~\cite{
Buzek90, STSATE85, Vidiella94}. It can be simply produced from the action of a
single mode squeeze operator on the state $\mid
p,0\rangle$~\cite{STSATE85}, where $p$ is the Bargmann
number. It also represents a linear combination from the number states with
coefficients chosen such that the photon counting probability distribution is
the binomial distribution with mean $M|p|$, where $0 < |p|<
1$.
\par The scope of this communication is to employ the spin squeezing and quantum 
entropy to  elucidate entanglement for two-atom system prepared initially in
Bell-state interacting with single cavity field prepared initially in a binomial
state. We introduce our Hamiltonian model and give exact expressions for the
full wave function of the atomic Bell-state and field systems, shedding light on
the important question of the relation between spin squeezing and quantum entropy
behaviors. We also examine the evolution of the quasiprobability distribution Husimi $Q$-function for the model under consideration. With the help of the quasiprobability distribution, one gains further insight into the mechanism of the interaction of the model subsystems. In the language of quantum information theory a definition of spin
squeezing is presented for this system. The utility of the definition is
illustrated by examining spin squeezing from the point of view of quatum
information theory for the present system. This analysis is applicable to any
quantum tripartite-state subject to spin squeezing with appropriate initial
conditions.
\section{model solution and the reduced density operator}
%===============================================================================
%\setlength{\baselineskip}{17pt}
\par Our system is as follows: A light field is initially in the binomial state (the
state which interpolates between the nonlinear coherent and the number states )
$$
|\psi_{F}(0)\rangle=|p,M\rangle=\sum_{n=0}^{\infty}~b_{n}|n\rangle,\eqno(1)
$$
where $|n\rangle$ is the eigenstate of number operator $\widehat{a}^{\dag}\widehat{a}=\widehat{n}$, $\widehat{n}|n\rangle=n|n\rangle$,  while, the coefficients $b_{n}$ are given by
$$
b_{n}=\sqrt{\binom{M}{n}~|p|^{n}(1-|p|)^{M-n}},\eqno(2)
$$
where $M>0$ (i.e., $M$ in general is any real positive number), $0 <|p|< 1$, 
interacts with two identical two-level atoms are initially in the following Bell
state
$$
|\psi_{A}^{\uparrow\uparrow\downarrow\downarrow}(0)\rangle=\gamma_{1}
|\uparrow\uparrow\rangle+\gamma_{4}|\downarrow\downarrow\rangle,\eqno(3)
$$
The binomial states have the properties
$$
|p,M\rangle=
\begin{cases}
 \mid M\rangle &|p|\rightarrow 1\cr
 \mid 0\rangle &|p|\rightarrow 0\cr
 \mid\alpha\rangle &|p|\rightarrow 0,~~~M\rightarrow \infty,~~~M|p|=\alpha^{2}.\cr
\end{cases}\eqno(4)
$$
For this system, the atoms-field interaction is governed by the ($N=2$)
Tavis-Cummings model(TCM)~\cite{TCM68}. The Tavis-Cummings model
(TCM)~\cite{TCM68} describes the simplest fundamental interaction between a
single mode of quantized electromagnetic field and a collection of $N$ atoms
under the usual two-level and rotating wave approximations when all of the atoms
couple {\it identically} to the field. The two-atom ($N=2$) TCM is governed by
the Hamiltonian
%\begin{widetext}
%\begin{center}
$$
\widehat{H}=\omega
\widehat{n}+\frac{1}{2}\omega_{0}(\widehat{\sigma}_{z}^{(1)}+\widehat{\sigma}_{z}^{(2)})+f(\chi,
\widehat{n})
%$$
%$$
+\lambda\sum_{j=1}^{2}\biggl(\widehat{\sigma}_{-}^{(j)}~\widehat{a}^{\dag}+\widehat{\sigma}_{+}^{(j)}~a\biggr)
,\eqno(5)
$$
%\end{center}
%\end{widetext}
where $f(\chi, \widehat{n})=\chi \widehat{n} (\widehat{n}-1)+2\sqrt{\chi}~
\widehat{n}$ represents the nonlinear term with,
$$f(\chi, \widehat{n})\mid n\rangle=\bigl(\chi
n(n-1)+2\sqrt{\chi} n\bigr)\mid n\rangle=f(\chi, n)\mid n\rangle.\eqno(6)
$$
We denote by $\chi$ the dispersive part of the third order susceptibility of the Kerr-like medium~\cite{MSAT07,MSAT109,MSAT209,MSAT309}. The parameter $\lambda$ represents the atoms-field coupling constant. The operators $\widehat{\sigma}_{\pm}^{(i)}$ and $\widehat{\sigma}_{z}^{(i)}$, ($i\in\{1,2\}$)
display a local $SU(2)$ algebra for the $i$-th atom in the 2D supspace spanned
by the ground (excited) state $|\downarrow\rangle$,($|\uparrow\rangle$) and obey the commutation
relations $[\widehat{\sigma}_{+}^{(i)},\widehat{\sigma}_{-}^{(i)}]=\widehat{\sigma}_{z}^{(i)}$,
($i\in\{1,2\}$), and $\widehat{a}$($\widehat{a}^{\dag}$) is bosonic annihilation (creation) operator
for the single mode field of frequency $\omega$. To make calculation to be more
clear and simple, we put
$$
\kappa\sum_{j=1}^{2}~\widehat{\sigma}_{s}^{(j)}=\widehat{J}_{s};~~~~~ s=z, -, +,\eqno(7)
$$
the parameter $\kappa=\frac{1}{2}, 1, 1$ if $s=z, -, +$, respectively.\\
The Hamiltonian (5), with the detuning parameter $\Delta=\omega_{0}-\omega$,  takes the form
$$
\widehat{H}=\widehat{H}_{0}+\widehat{H}_{INT}=\omega\bigl(\widehat{n}+\widehat{J}_{z}\bigr)+\Delta~\widehat{J}_{z}+f(\chi,
\widehat{n})
%$$
%$$
+\lambda~\bigl(\widehat{J}_{-}~\widehat{a}^{\dag}+\widehat{J}_{+}~\widehat{a}\bigr),\eqno(8)
$$
Consider, at $t=0$, the two atoms are in the Bell state, Eq. (3). The initial
state of the system is a decoupled pure state, and the sate vector can be
written as
$$
|\psi_{AF}(0)\rangle=\sum_{n=0}^{\infty}b_{n}\bigl[\gamma_{1}|\uparrow\uparrow,
n\rangle+\gamma_{4}|\downarrow\downarrow,n\rangle\bigr],\eqno(9)
$$
As the time goes, the evolution of the system in the interaction picture can be
obtained by  solving the Schr\"{o}dinger equation
$$
i\frac{d}{dt}|\psi_{AF}(t)\rangle=H|\psi_{AF}(t)\rangle,\eqno(10)
$$ 
and using the condition Eq. (9) to obtain the time-dependent wave function in the
form
$$
|\psi_{AF}
(t)\rangle=|U\rangle~|1\rangle+|R\rangle~|2\rangle+|S\rangle~|3\rangle+|T\rangle
~|4\rangle,\eqno(11)
$$
where
$$
|U\rangle=\sum_{n=0}^{\infty}~b_{n}\biggl[A_{n}(t)|n\rangle+H_{n-2}
(t)|n-2\rangle\biggr],\eqno(12)
$$
$$
|R\rangle=\sum_{n=0}^{\infty}~b_{n}\biggl[B_{n+1}(t)|n+1\rangle+G_{n-1}
(t)|n-1\rangle\biggr]=|S\rangle,\eqno(13)
$$
and
$$
|T\rangle=\sum_{n=0}^{\infty}~b_{n}\biggl[D_{n+2}(t)|n+2\rangle+E_{n}
(t)|n\rangle\biggr],\eqno(14)
$$
where we used the notations
$$
|1\rangle=|\uparrow\uparrow\rangle,~~~~|2\rangle=|\uparrow\downarrow\rangle~~~~|3\rangle=|\downarrow\uparrow\rangle~~~~~|4\rangle=|\downarrow\downarrow\rangle.\eqno(15)
$$
the group of the complex amplitudes $A_{n}(t)$, $D_{n+2}(t)$, $B_{n+1}(t)$ and $C_{n+1}(t)$ are given, respectively by
$$
A_{n}(t)=\frac{1}{2\Gamma_{1}\Gamma_{2}}\sum_{k=0}^{2}
D^{k}_{n+2}(0)[\eta^{2}_{k}-2\Gamma_{2}^{2}+(\alpha_{2}+\alpha_{3})\eta_{k}
+\alpha_{2}\alpha_{3}]e^{i\eta_{k} t},\eqno(16)
$$
$$ 
B_{n+1}(t)=-\frac{1}{2\Gamma_{2}}\sum_{k=0}^{2}D^{k}_{n+2}(0)(\eta_{k}+\alpha_{3
})e^{i\eta_{k} t}=C_{n+1}(t),\eqno(17)
$$
$$
D_{n+2}(t)=\sum_{k=0}^{2} D^{k}_{n+2}(0)e^{i\eta_{k} t},\eqno(18)
$$
with
$$
\alpha_{1}=\Delta+f(\chi, n),~~~ \alpha_{2}=f(\chi, n+1),~~~ \alpha_{3}=-\Delta+f(\chi, n+2)\eqno(19)
$$
$$
\Gamma_{1}=\lambda \sqrt{n+1},~~~\Gamma_{2}=\lambda \sqrt{n+2},\eqno(20)
$$
where
$$ 
\eta_{k}=-\frac{X_{1}}{3}+\frac{2}{3}\biggl(\sqrt{X_{1}^{2}-3X_{2}}\biggr)
\cos 
(\theta^{k}),\eqno(21)
$$ 
with
 $$ 
\theta^{k}=\biggl(
 \frac{1}{3}\cos^{-1}
\biggl[
\frac{9X_{1}X_{2}-2X_{1}^{3}-27X_{3}}{2(X_{1}^{2}-3X_{2})^{\frac{3}{2}}}
\biggr]+\frac{2k\pi}{3}\biggr),
 k=0,1,2,\eqno(22)
$$
and
$$
X_{1}=\alpha_{1}+\alpha_{2}+\alpha_{3},~~~X_{2}=\alpha_{2}\alpha_{3}+\alpha_{1}(\alpha_{2}+\alpha_{3})-2(\Gamma_{1}^{2}
+\Gamma_{2}^{2}),~~~ X_{3}=\alpha_{1}\alpha_{2}\alpha_{3}-2(\alpha_{1}\Gamma_{2}^{2}+\alpha_{3}
\Gamma_{1}^{2}), \eqno(23)
$$
where the complex coefficients $D^{k}_{n+2}(0)$, $k=0,1,2$ are given by
$$
D^{k}_{n+2}(0)=\frac{2\gamma_{1}\Gamma_{1}\Gamma_{2}}{\eta_{kr}\eta_{ks}}~~~~,k,
r,s=0,1,2;~~~~ k\neq r\neq s \eqno(24)
$$
If in Eqs. (16-24), we let $\gamma_{1}\rightarrow \gamma_{4}$ and 
$$
\alpha_{1}=-\Delta+f(\chi, n),~~~\alpha_{2}=f(\chi, n-1),~~~ \alpha_{3}=\Delta+f(\chi, n-2),\eqno(25)
$$
$$
\Gamma_{1}=\lambda \sqrt{n},~~~\Gamma_{2}=\lambda \sqrt{n-1},\eqno(26)
$$
and replacing the group of complex amplitudes $A_{n}(t)$, $D_{n+2}(t)$, $B_{n+1}(t)$ and $C_{n+1}(t)$ by the other group of complex amplitudes $E_{n}(t)$, $F_{n-1}(t)$, $G_{n-1}(t)$, $H_{n-2}(t)$,  we obtain easily the last group, respectively.\\
The reduced density operator of the subsystem is given by
$$
\rho(t)_{A(F)}=\mathbf{Tr}_{F(A)}\rho(t)_{AF}=\mathbf{Tr}_{F(A)}|\psi_{AF}
(t)\rangle\langle\psi_{AF}(t)|.\eqno(27)
$$
Then the reduced atomic density operator in matrix form is given by
\begin{displaymath}
\mathbf{\rho_{A}} = \left( \begin{array}{cccc}
\rho_{A}^{11}~&~\rho_{A}^{12}~&~\rho_{A}^{13}~&~\rho_{A}^{14}\\
\\
\rho_{A}^{21}~&~\rho_{A}^{22}~&~\rho_{A}^{23}~&~\rho_{A}^{24}\\
\\
\rho_{A}^{31}~&~\rho_{A}^{32}~&~\rho_{A}^{33}~&~\rho_{A}^{34}\\
\\
\rho_{A}^{41}~&~\rho_{A}^{42}~&~\rho_{A}^{43}~&~\rho_{A}^{44}\\
\end{array} \right),\eqno(28)
\end{displaymath}
where the elements $\rho_{A}^{sr}$, $s,r\in\{1,2,3,4\}$ are given by
$$
\rho_{A}^{11}=\langle
U|U\rangle=\sum_{n=0}^{\infty}\biggl\{|b_{n}|^{2}\biggl(|A_{n}|^{2}+|H_{n-2}
|^{2}\biggr)
%$$
%$$
+b_{n}b_{n+2}^{\ast}A_{n}H_{n}^{\ast}+b_{n}b_{n-2}^{\ast}A_{n-2}^{\ast}H_{n-2}
\biggr\},\eqno(29)
$$
$$
\rho_{A}^{22}=\langle
R|R\rangle=\sum_{n=0}^{\infty}\biggl\{|b_{n}|^{2}\biggl(|B_{n+1}|^{2}+|G_{
n-1}|^{2}\biggr)
%$$
%$$
+b_{n}b_{n+2}^{\ast}B_{n+1}G_{n+1}^{\ast}+b_{n}b_{n-2}^{\ast}B_{n-1}^{\ast}G_{n-1}
\biggr\}=\langle
S|S\rangle=\rho_{A}^{33},\eqno(30)
$$
$$
\rho_{A}^{44}=\langle
T|T\rangle=\sum_{n=0}^{\infty}\biggl\{|b_{n}|^{2}\biggl(|D_{n+2}|^{2}+|E_{n}
|^{2}\biggr)
%$$
%$$
+b_{n}b_{n+2}^{\ast}D_{n+2}E_{n+2}^{\ast}+b_{n}b_{n-2}^{\ast}D_{n}^{\ast}E_{n}
\biggr\},\eqno(31)
$$
$$
\rho_{A}^{12}=\langle
R|U\rangle=\sum_{n=0}^{\infty}\biggl\{b_{n}b_{n-1}^{\ast}A_{n}B_{n}^{\ast}
%$$
%$$
+b_{n}b_{n-3}^{\ast}B_{n-2}^{\ast}H_{n-2}
$$
$$
+b_{n}b_{n+1}^{\ast}A_{n}G_{n}^{\ast}
%$$
%$$
+b_{n}b_{n-1}^{\ast}G_{n-2}^{\ast}H_{n-2}\biggr\}=(\rho_{21})^{\ast}=\langle
S|U\rangle=\rho_{A}^{13},\eqno(32)
$$
$$
\rho_{A}^{14}=\langle
T|U\rangle=\sum_{n=0}^{\infty}\biggl\{b_{n}b_{n-2}^{\ast}A_{n}D_{n}^{\ast}
%$$
%$$
+b_{n}b_{n-4}^{\ast}D_{n-2}^{\ast}H_{n-2}
$$
$$
+|b_{n}|^{2}A_{n}E_{n}^{\ast}
%$$
%$$
+b_{n}b_{n-2}^{\ast}E_{n-2}^{\ast}H_{n-2}\biggr\}=(\rho_{41})^{\ast},\eqno(33)
$$
$$
\rho_{A}^{23}=\langle
S|R\rangle=\sum_{n=0}^{\infty}\biggl\{|b_{n}|^{2}|B_{n+1}|^{2}
%$$
%$$
+b_{n}b_{n-2}^{\ast}C_{n-1}^{\ast}G_{n-1}
$$
$$
+b_{n}b_{n+2}^{\ast}B_{n+1}F_{n+1}^{\ast}
%$$
%$$
+|b_{n}|^{2}|F_{n-1}|^{2}\biggr\}=\langle
R|S\rangle=\rho_{32},\eqno(34)
$$
$$
\rho_{A}^{24}=\langle
T|R\rangle=\sum_{n=0}^{\infty}\biggl\{b_{n}b_{n-1}^{\ast}B_{n+1}D_{n+1}^{\ast}
%$$
%$$
+b_{n}b_{n-3}^{\ast}D_{n-1}^{\ast}G_{n-1}
$$
$$
+b_{n}b_{n+1}^{\ast}B_{n+1}E_{n+1}^{\ast}
%$$
%$$
+b_{n}b_{n-1}^{\ast}E_{n-1}^{\ast}G_{n-1}\biggr\}=(\rho_{42})^{\ast}=\rho_{A}^{34},\eqno(35)
$$
\section{Spin Squeezing}
Spin squeezing phenomenon reflects the reduced quantum
fluctuations in one of the field quadratures at the expense of the other
corresponding stretched quadrature.
In the literature,~\cite{KIUE93,WBI94,PZM02, SDCZ01,ZPP91,WIZO81}, there are
several definitions of spin squeezing  and which one is the best is still an
unsolved issue. Squeezing or reduction of quantum fluctuations, for arbitrary
operators $A$ and $B$ which obey the commutation relation $[A,B]=C$, is the
product of the uncertainties in determining their expectation values as
follows~\cite{WIZO81}:
$$
\Delta A \Delta B\geq\frac{1}{2}|\langle C\rangle|,\eqno(36)
$$
where $(\Delta A)^2=\langle A^2\rangle-\langle A\rangle^2$ and $(\Delta
B)^2=\langle B^2\rangle-\langle B\rangle^2$. \\
In this work spin squeezing parameters are based on angular momentum commutation
relations. From the commutation relation $[J_{x},J_{y}]=i J_{z}$ the uncertainty
relation between different componenets of the angular momentum given by
$$
\Delta J_{x} \Delta J_{y}\geq\frac{1}{2}|\langle J_{z}\rangle|,\eqno(37)
$$
where
$$
J_{x}=\frac{1}{2}(J_{+}+J_{-}),\eqno(38)
$$
$$
J_{y}=\frac{1}{2i}(J_{+}-J_{-}),\eqno(39)
$$
where the operators $J_{+}$, $J_{-}$ and $J_{z}$ are given by (6).
Without violating Heisenberg's uncertainty relation, it is possible to
redistribute the uncertainty unevenly between $J_{x}$ and $J_{y}$, so that a
measurement of either $J_{x}$ or $J_{y}$ becomes more precise than the standard
quantum limit $\sqrt{|\langle J_{z}\rangle|/2}$. States with this property are
called spin squeezed states in analogy with the squeezed states of a harmonic
oscillator. Consequently, the two squeezing parameters can be written as
$$
F_{1}=(\Delta J_{x})^2-\frac{1}{2}|\langle
J_{z}\rangle|=\frac{1}{2}\biggl(1-\frac{1}{2}\biggl\langle
(J_{+}+J_{-})\biggr\rangle^{2}-|\langle J_{z}\rangle|\biggr),\eqno(40)
$$
$$
F_{2}=(\Delta J_{y})^2-\frac{1}{2}|\langle
J_{z}\rangle|=\frac{1}{2}\biggl(1-\frac{1}{2i}\biggl\langle
(J_{+}-J_{-})\biggr\rangle^{2}-|\langle J_{z}\rangle|\biggr),\eqno(41)
$$
If the parameter $F_{1}$ ($F_{2}$) satisfies the condition $F_{1}<0$
($F_{2}<0$), the fluctuation in the component $J_{x}$ ($J_{y}$) is said to be
squeezed.\\
Using the wave function (11), we can easily compute the following time-dependent
expectation values of the operators $J_{-}$, $J_{+}$ and $J_{z}$, respectively
in the forms
$$
\langle
J_{-}\rangle=2(\rho_{A}^{12}+\rho_{A}^{24})=\langle
J_{+}\rangle^{\ast}
,\eqno(42)
$$
$$
\langle J_{z}\rangle=\rho_{A}^{11}-\rho_{A}^{44}
,\eqno(43)
$$
\section{Quantum Entropy}
Quantum entropy, as a natural generalization of Boltzmann classical entropy, was
proposed by von Neumann~\cite{NEUMANN27}. It has been applied, in particular, as
a measure of quantum entanglement, quantum decoherence, quantum optical
correlations, purity of states, quantum noise or accessible information in
quantum measurement (the capacity of the quantum channel). Entropy is related to the density matrix, which provides a
complete statistical description of the system.
Since we have assumed that the two two-level atoms and the single-mode binomial
field are initially in a disentangled pure state, the total entropy of the
system is zero. In terms of Araki $\&$ Lieb inequality of the
entropy~\cite{ARLE70}
$$
|S_{A}(t)-S_{F}(t)|\leq S_{AF}(t)\leq |S_{A}(t)+S_{F}(t)|,\eqno(44)
$$
we can find that the reduced entropies of the two subsystems are identical,
namely, $S_{A}(t)=S_{F}(t)$. Here, the field entropy can be obtained by
operating the atomic entropy. The quantum field entropy can be defined as follows~\cite{VPRK97}
$$
S(\rho_{A})=-\sum_{s=1}^{4}~\Pi_{A}^{(s)}\ln \Pi_{A}^{(s)},\eqno(45)
$$
where $\Pi_{A}^{(s)}$, ($s=1,2,3,4$) is the eigenvalue of the reduced density
matrix, Eq. (27), and can be represented by the roots of forth order algebraic equation
$$
c_{0}+c_{1}\Pi_{A}+c_{2}\Pi_{A}^{2}+c_{3}\Pi_{A}^{3}+\Pi_{A}^{4}=0,\eqno(46)
$$
with the coefficients, $c$'s, are given by
$$
c_{3}=-\rho_{11}-\rho_{22}-\rho_{33}-\rho_{44},\eqno(47)
$$
$$
c_{2}=-|\rho_{41}|^{2}-2|\rho_{42}|^{2}-2|\rho_{12}|^{2}-|\rho_{23}|^{2}
%$$
%$$
+2\rho_{22}(\rho_{11}+\rho_{44})+\rho_{22}^{2}+\rho_{11}\rho_{44},\eqno(48)
$$
$$
c_{1}=2|\rho_{14}|^{2}\rho_{22}+2|\rho_{24}|^{2}(\rho_{11}+\rho_{22})+2|\rho_{12
}|^{2}(\rho_{22}+\rho_{44})
%$$
%$$
-\rho_{22}^{2}(\rho_{11}+\rho_{44})
$$
$$
-2\rho_{11}\rho_{22}\rho_{44}-\Re(\rho_{23})(|\rho_{42}|^{2}+|\rho_{12}|^{2})
%$$
%$$
-2\Re(\rho_{41}\rho_{12}\rho_{24})+|\rho_{32}|^{2}(\rho_{11}+\rho_{44}),
\eqno(49)
$$
$$
c_{0}=|\rho_{41}|^{2}|\rho_{23}|^{2}-\rho_{22}\rho_{33}|\rho_{41}|^{2}-\rho_{11}
\rho_{33}|\rho_{42}|^{2}
%$$
%$$
-\rho_{44}\rho_{22}|\rho_{12}|^{2}-\rho_{11}\rho_{44}|\rho_{23}|^{2}-\rho_{33}
\rho_{44}|\rho_{12}|^{2}
$$
$$
+\rho_{11}\rho_{22}\rho_{33}\rho_{44}
-\rho_{11}\rho_{22}|\rho_{24}|^{2}
%$$
%$$
+(\rho_{11}|\rho_{24}|^{2}+\rho_{44}|\rho_{12}|^{2})\Re(\rho_{23})
%$$
%$$
+(\rho_{22}+\rho_{33}-\Re(\rho_{23}))\Re(\rho_{41}\rho_{12}\rho_{24}),\eqno(50)
$$
where we used the notations
$$
\rho_{22}=\rho_{33},~~~~~\rho_{12}= \rho_{13},~~~~~\rho_{24}=
\rho_{34}.\eqno(51)
$$
The eigenvalues $\Pi_{A}^{(s)}$, ($s=1,2,3,4$) are given, respectively by
$$
\Pi_{A}^{(s)}=\frac{U_{s}+(-1)^{s}V_{s+1}}{2};~~~~ s=1,2\eqno(52)
$$
$$
\Pi_{A}^{(s)}=\frac{U_{s}+(-1)^{s+1}V_{s+2}}{2};~~~~ s=3,4\eqno(53)
$$
where
$$
U_{s}=-\frac{c_{3}}{2}+(-)^{s}f, V_{s}=\sqrt{z_{3}+(-1)^{s}z_{4}},\eqno(54)
$$
with
$$
z_{3}=-2~c_{2} + \frac{3~c_{3}^{2}}{4} -f^{2}, z_{4}=\frac{8 c_{1} - 4 c_{2}
c_{3} + c_{3}^{3}}{4~f}\eqno(55)
$$
\section{Husimi $Q$-function}
For measuring the quantum state of the radiation field, balanced homodyning has become a well established method, it directly measures phase sensitive quadrature distributions. The homodyne measurement of an electromagnetic field gives all possible linear combinations of the field quadratures. The average of the random outcomes of the measurement is connected with the marginal distribution of any quasi-probability used in quantum optics. It has been shown from earlier studies~\cite{EWIGNER32, ZWIGNer32, KAGL69,
HICOSCWG84} that the quasi-probability
functions $W$-, (Husimi) $Q$-, and (Glauber-Sudershan) $P$-function, are important for the statistical description of a microscopic system
and provide insight into the nonclassical features of the radiation fields.
Therefore, we devote the present section to concentrate on one of these functions, that is the Husimi $Q$-function which has the nice property of being always positive and further advantage of being readily measurable by quantum tomographic techniques~\cite{BOTAWA98, MANTOM97}. In fact, Husimi $Q$-function is not only a convenient tool to calculate the expectation values of anti-normally ordered products of operators, but also to give some insight into the mechanism of interaction for the model under consideration. The relation between the phase-space measurement; Husimi $Q$-function; and the classical information-theoretic entropy associated with quantum fields was introduced by Wehrl~\cite{WEHRL79}. However, on expanding the von Neumann quantum entropy in a power series of classical entropies, it was shown explicity~\cite{PEKRPELUSZ86} that the first term of this expansion is the Wehrl entropy. Thus, Husimi $Q$-function can be related to the von Neumann quantum entropy in different approaches~\cite{WEHRL79, PEKRPELUSZ86, FAGU06, CAALCARA09, HUFAN09, MIMAWA00, MIWAIM01, BERETA84}.\\
The Husimi $Q$-function can be given in the form as
~\cite{HICOSCWG84, HUSIMI40, FuSOLO001}
$$
Q(\alpha)=\frac{\langle \alpha\mid\rho_{F}\mid\alpha\rangle}{\pi},\eqno(56)
$$
where $\rho_{F}$ is the reduced density operator of the cavity field given by (27). The state $\mid\alpha\rangle$ represents the well-known coherent state with amplitude $\alpha=X+i Y$. Inserting $\rho_{F}$ into Eq. (56), we
can easily obtain the Husimi $Q$-function of the cavity field
$$
Q(\alpha)=\frac{1}{\pi}(\mid\langle\alpha\mid
U\rangle\mid^{2}+\mid\langle\alpha\mid T\rangle\mid^{2})\eqno(57)
$$
where
$$
\langle\alpha\mid
U\rangle=e^{-\mid\alpha\mid^{2}/2}\sum_{n=0}^{\infty}\biggl[b_{n}\frac{\alpha^{
\ast n}}{\sqrt{n!}}A_{n}(t)+b_{n+2}\frac{\alpha^{\ast
n}}{\sqrt{n!}}H_{n}(t)\biggr]\eqno(58)
$$
and
$$
\langle\alpha\mid
T\rangle=e^{-\mid\alpha\mid^{2}/2}\sum_{n=0}^{\infty}b_{n}\biggl[\frac{\alpha^{
\ast n}}{\sqrt{n!}}E_{n}(t)+\frac{\alpha^{\ast n+2}}{\sqrt{(n+2)!}}D_{n+2}(t)\biggr]\eqno(59)
$$
%\newpage
%
%
\section{Discussion of the numerical results}
Using different sets of parameters for the initial state of the system we
calculate numerically the quantum entropy $S_{A}$, spin squeezing parameters
$F_{1}$ and $F_{2}$ and atomic population $\langle \sigma_{z}\rangle$ as a
reference function. All results are plotted as functions of the Rabi angle
$\lambda t$. For each set of parameters four pictures are displayed. The pictures (a and b) show, respectively, squeezing parameters $F_{1}$ and
$F_{2}$, while the pictures (c and d) show the quantum entropy
$S_{A}$ and the atomic population $\langle \sigma_{z}\rangle$. For
all our plots we set the Bell-state parameters $\gamma_{1}=\frac{1}{\sqrt{2}}$
and $\gamma_{4}=i\gamma_{1}$. In figures 1 and  2 we have plotted the above
mentioned functions with $|p|=0.9$, $\chi/\lambda=0$ and various values of the
detuning parameter $\Delta/\lambda$. From these figures we can easily notice
that, just after the onset of interaction these functions fluctuate for short
period of time. This short period of revival is followed by a long period of
collapse. The period of revival starts longer for one period of time with high
amplitude to become wider with smaller amplitude as time goes. 
This is because
the width and heights of the revivals belonging to the different series of
eigenvalues are different.
Furthermore, as we increase the detuning parameter $\Delta/\lambda$ from its
value $\Delta/\lambda=0$ (resonance case), the overlap of revivals noticeably
decreases with the increase in $\Delta/\lambda$. Also, the periods of
oscillations within the revival decrease with the increase in the detuning
parameter $\Delta/\lambda$. 
For the population, $\langle \sigma_{z}\rangle$, the
period of revival depends on the average number $M|p|$ of photons whereas the
time of collapse depends on the dispersion, $M|p|(1-|p|)$, in the photon number
distribution~\cite{JoshPur87}. 
\begin{figure}[h]
%\noindent
\begin{center}
\includegraphics[width=.9\linewidth]
{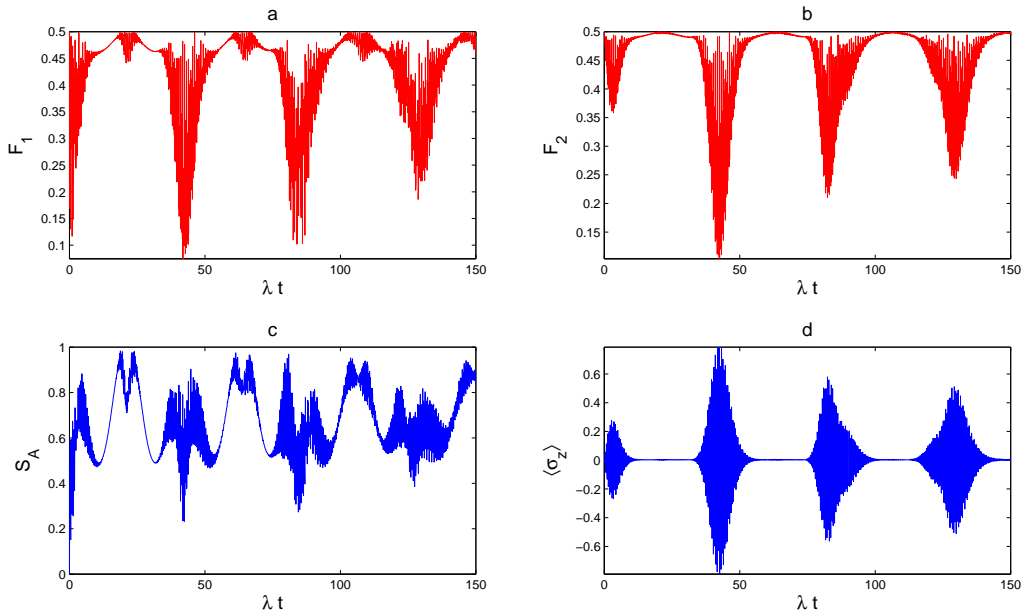}
\vspace{-.7cm}
\caption{Spin squeezing parameters $F_{1}$ (a), $F_{2}$ (b), atomic
entropy $S_{A}$ (c), and atomic inversion (d) with $|p|=0.9$,
$M=50$,  $\chi/\lambda=0$ and $\Delta/\lambda=0$}
\end{center}
\end{figure}
Moreover, from these figures we can see that spin
squeezing parameters $F_{1}$ and $F_{2}$ experience collapse and revival where
atomic population exhibits collape and revival as time going.
%
%\newpage
% 
%
When we turn our attention to the role that spin squeezing parametres play to
discover entanglement properties, we can realize that the behaviors of both
squeezing parameter $F_{1}$ and atomic entropy $S_{A}$ are equivalent, i.e.,
entanglemet implies spin squeezing and vice versa.
This can be understood as follows: quantum entropy $S_{A}$ oscillates when
$F_{1}$ exhibits oscillations with same periods of time. 
\begin{figure}[h]
\begin{center}
\includegraphics[width=0.9\linewidth]
{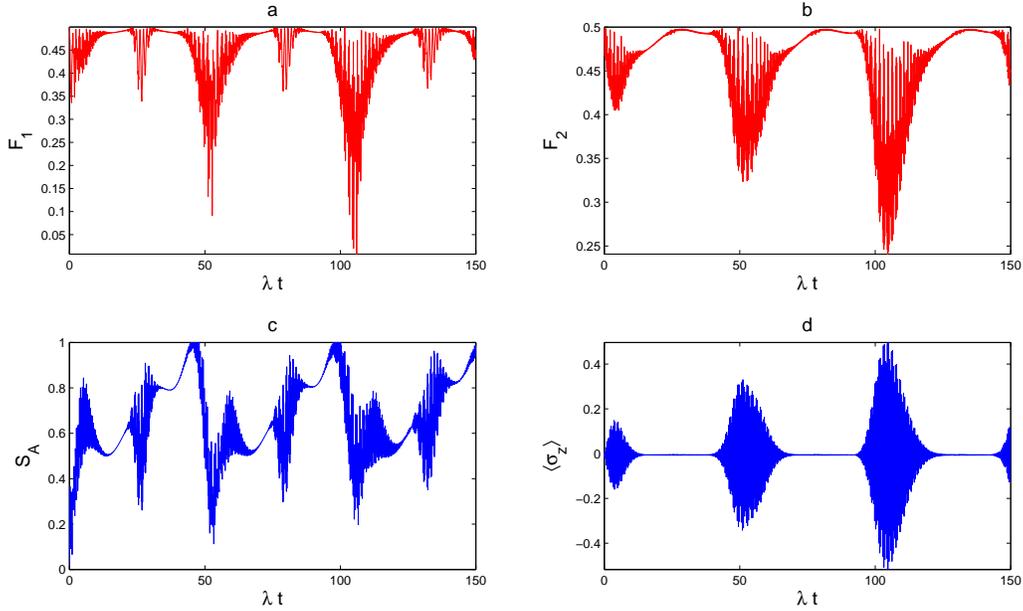}
\vspace{-0.5cm}
\caption{ The same as Fig. (1) but for $\Delta/\lambda=10$}
\end{center}
\end{figure}
\begin{figure}[h]
\begin{center}
\includegraphics[width=0.9\linewidth]
{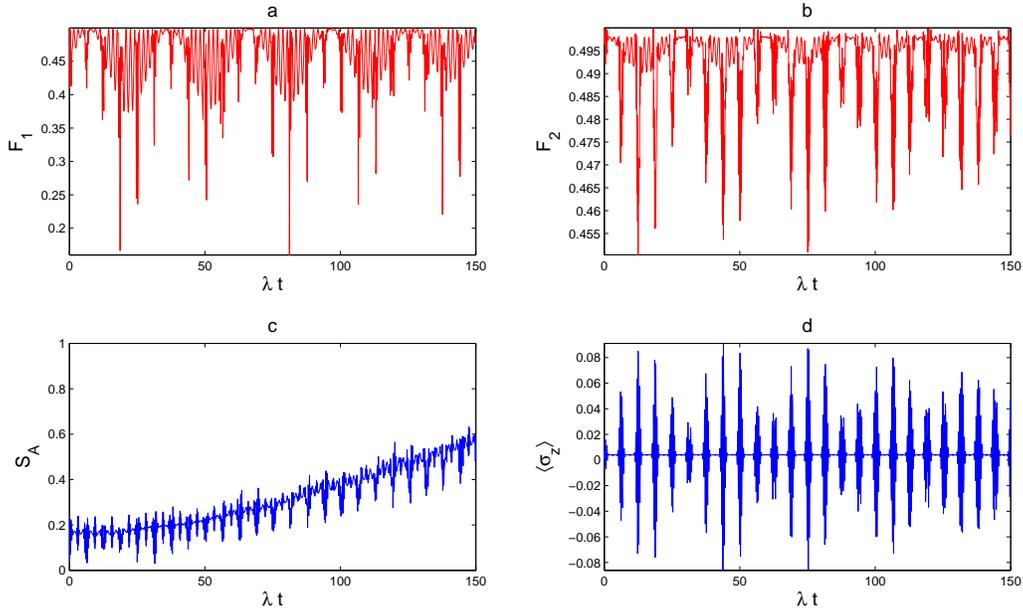}
\vspace{-0.5cm}
\caption{The same as Fig. (1) but for $\chi/\lambda=0.5$ }
\end{center}
\end{figure}
Furthermore, the
function $S_{A}$ goes to its maximum when $F_{1}$ shows oscillations around very
small value (between 0.45 and 0.5) of its maximum and when $F_{2}$ shows
collapse equal to its maximum, while $S_{A}$ reaches its minimum when squeezing
occurs.
This behavior occurs periodically for both $S_{A}$ and $F_{1}$. This means that,
on on-resonance atomic-system-field interaction, we can, with full success,
understand entanglement dynamics from the dynamic of spin squeezing parameters
$F_{1}$ and $F_{2}$ and vice versa.\\
\begin{figure}[h]
\begin{center}
\includegraphics[width=0.9\linewidth]
{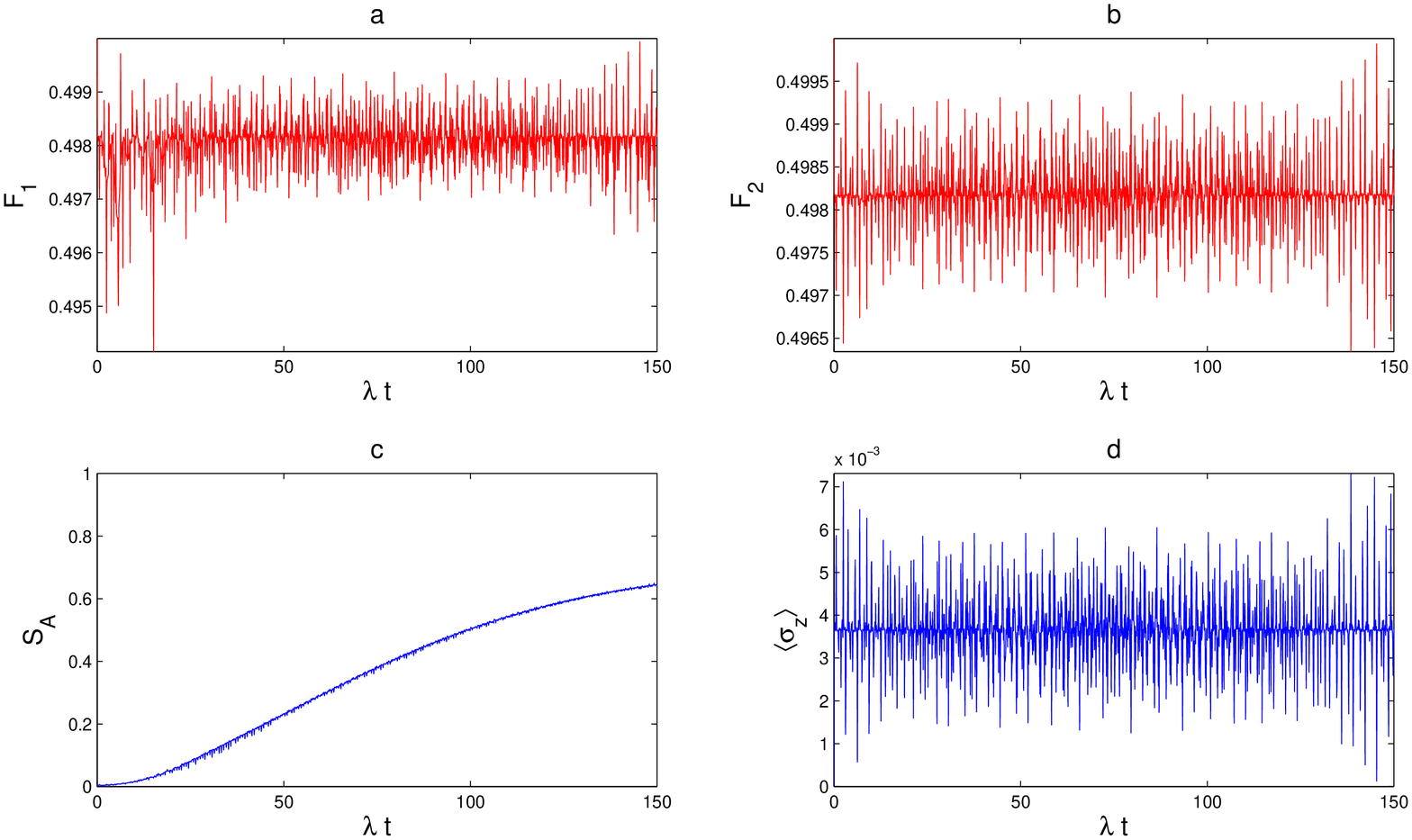}
\vspace{-0.5cm}
\caption{ The same as Fig. (1) but for $\chi/\lambda=5.0$ }
\end{center}
\end{figure}
Let us now come to the case of off-resonance interaction between the atomic
system and the cavity field. In this case the same general behavior (with
periods shift to right when $\Delta/\lambda=10$) is noticed. Additionally, the
oscillations become more dense with reduced maximum of $S_{A}$ corresponding to
the increase of the oscillation interval of $F_{1}$ (between 0.4 and 0.5 and
become longer as $\Delta/\lambda$ increase) and when some intervals of collapse
begin to appear.\\
The surprising and very interesting is the effect of the nonlinear medium
individually and in the presence of the detuning parameter $\Delta/\lambda$.
\begin{figure}[h]
\begin{center}
\includegraphics[width=0.9\linewidth]
{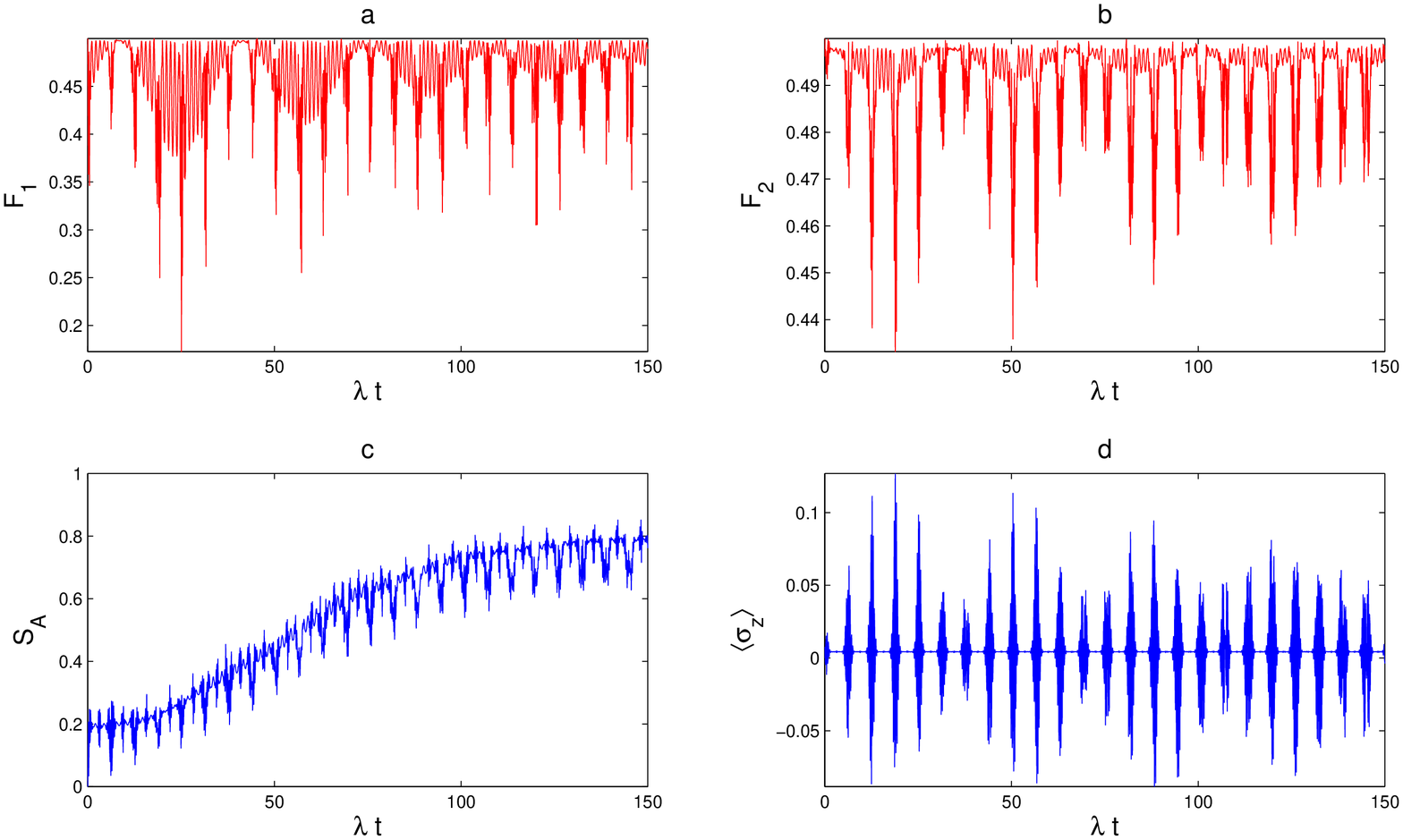}
\vspace{-0.5cm}
\caption{The same as Fig. (3) but for $\Delta/\lambda=5.0$ }
\end{center}
\end{figure}
To
examine the effect of these mentioned parameters, we recall figures (3-5). These
figures have been pictured by the setting of different values of the parameter
$\chi/\lambda$ individually and in the presence of the detuning parameter
$\Delta/\lambda$. It is easy to see the change in figures shape where the
standard behavior was changed completely. For a weak Kerr medium such that
$\chi/\lambda=0.5$, our reference function, $\langle\sigma_{z}\rangle$, shows
behavior similar to the modified Jayned-Cumming model with Kerr
medium~\cite{ETCU63,JOSHPUR92} accompanied with reduced amplitude of
oscillations. 
\begin{figure}[h]
\begin{center}
\includegraphics[width=0.9\linewidth]
{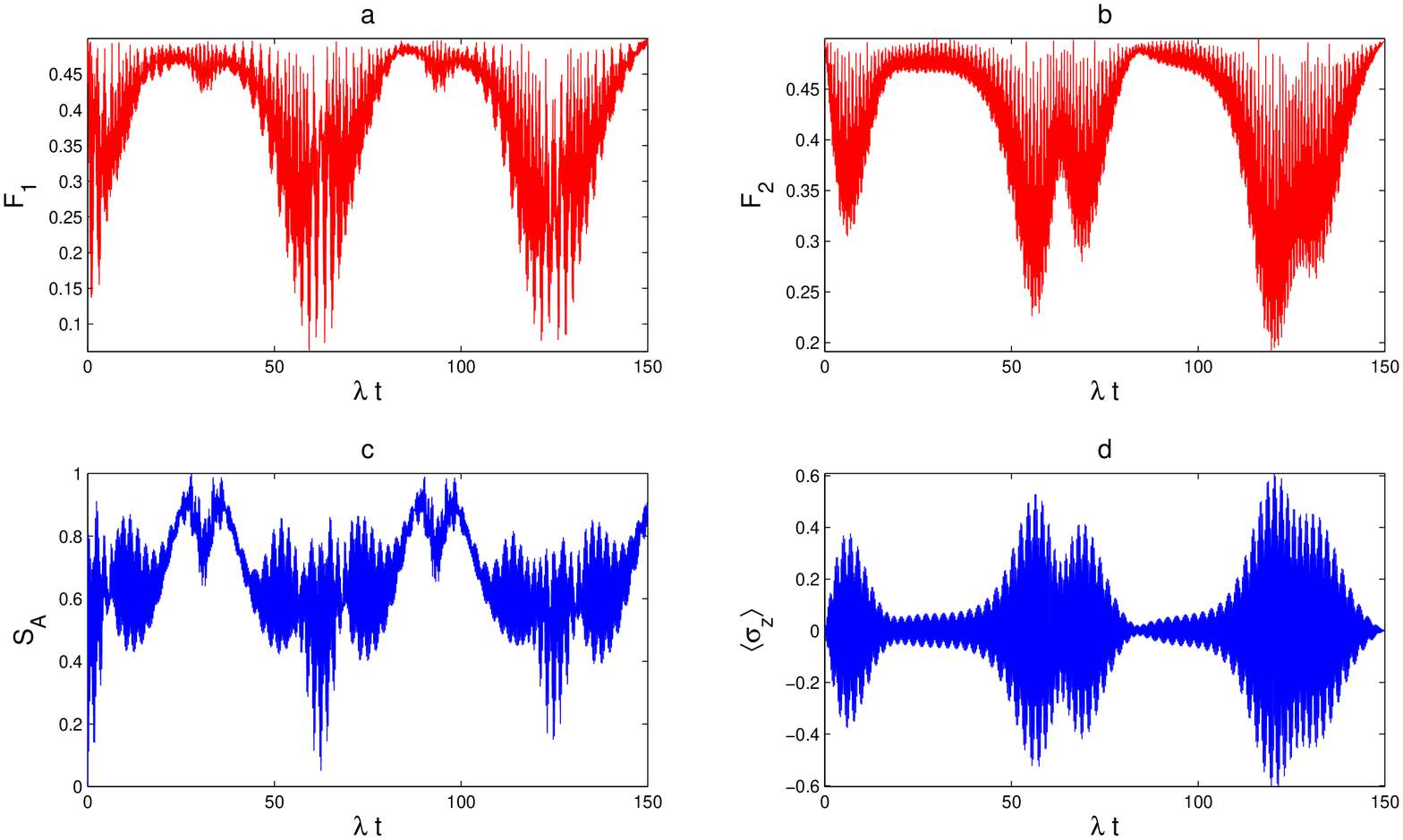}
\vspace{-0.5cm}
\caption{The same as Fig. (1) but for $|p|=0.98$, $M=100$}
\end{center}
\end{figure}
Furthermore, the population $\langle\sigma_{z}\rangle$ and spin squeezing parameter $F_{2}$ oscillate periodically with fixed periods are equivalent but $F_{1}$ does not. 
A quick look at the squeezing parameter $F_{1}$ one can realize easily that it oscillates rapidly. The oscillations overlap for a period of time (except for some instances) to become dense to show periodically wave packets of Gaussian envelope with amplitude decreases as time develops. The more the Gaussian-packet-envelope amplitude decreases the more the entropy increases, i.e., stronger entanglement can be showed, see figure 3. This behavior becomes more clear when we consider the detuning parameter in our numerical computations, see figure 5. 
\vspace{-0.5cm}
\begin{center}
\begin{figure}[h]
\includegraphics[width=0.9\linewidth]
{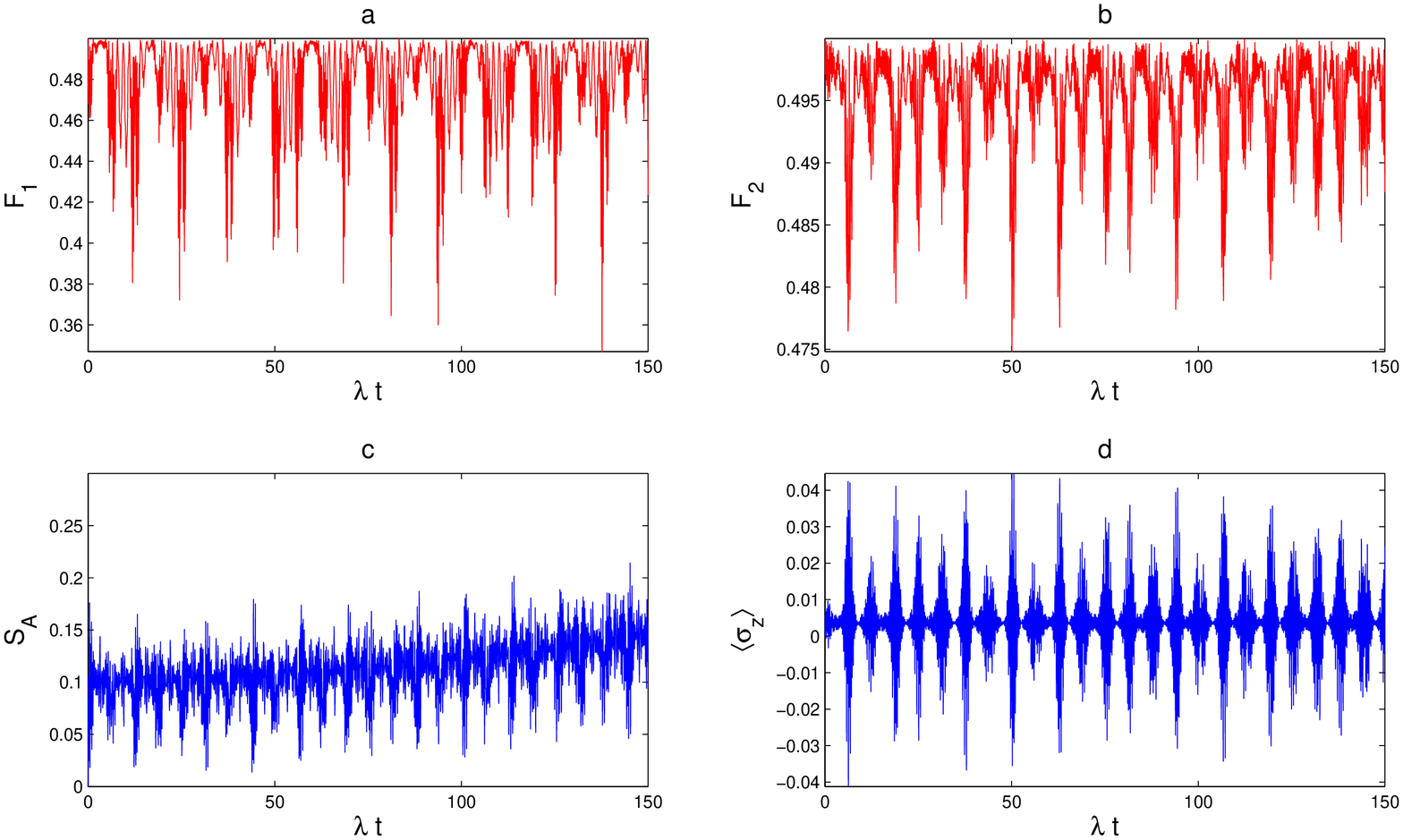}
\vspace{-0.5cm}
\caption{The same as Fig. (6) but for $\chi/\lambda=0.5$  }
\end{figure}
\end{center}
%
%\vspace{-0.7cm}
%
Furthermore, when the nonlinear medium becomes stronger, $\langle \sigma_{z}\rangle$, $F_{1}$ and $F_{2}$ show chaotic behavior with no indications of revivals or any other regular structure. This is
accompanied with change in the entropy maximum from slow to rapid increase as
$\chi/\lambda$ increases with time develops (see Fig. 4). This behavior is
dominant without or with high values of $\Delta/\lambda$.
\\
With the increase of $|p|$ and $M$, such that $|p|=0.98$, and $M=100$, i.e.; on
increasing the average number $M|p|$ of photon, the oscillations of squeezing
parametres $F_{1}$, $F_{2}$, $\langle \sigma_{z}\rangle$ and quantum entropy
$S_{A}$ becomes more dense. This means that every two neighbor overlapping
revivals start to overlap again. This is because with larger mean photon $M|p|$,
these functions have bigger values with time evolution, which causes dense
oscillations of the cavity field parametres. In other words, there are more
revival series since more eigenvalues can be found in this model. However, the same behavior, we saw before for $|p|=0.9$, is seen again except
for the envelope width becomes wider which resulted in fewer packets of
oscillation appear in the same period of time, see figure 6 and 7. It is
worth to note that each revival series of oscillations corresponds to a beat
frequency. Moreover, on weak Kerr medium the relation between the atomic entropy
and spin squeezing parametres seems more complicated. In this case all of them
oscillate with no indications of revivals or any other regular structure where
the Gaussian envelope completely disappears. The effect of
strong Kerr medium individually and with the coexistence of small detuning
exhibits behavior similar to that when $|p|=0.9$.
\par At the end we are going to focus our attention on the representation of the field in phase space which provides some aspects of the field dynamics. Figure 8 shows mesh plots (left) and the corresponding contour plots (right) of the Husimi $Q$-function in the complex $\alpha$-plane $X=Re(\alpha)$, $Y=Im(\alpha)$ for the Rabi angle $\lambda t=\pi/4$ and different values of Kerr parameter $\chi/\lambda$, while all other parameters are kept without change as in Fig. 1. From figure 8a, it is clear that the state of the field is a squeezed state, since the Husimi $Q$-function has different widths in the $X$ and $Y$ directions. On the other hand, the squeezing is generated by the nonlinearity inherent to the system by the binomial state. This can be explained in terms of a superposition of different numbers of states of different phases, creating a deviation from the classical phase.
It is well known that the squeezed states is a general class of the minimum-uncertainty states~\cite{SACHNEBU03}. Bearing in mind that the nonlinearity of the binomial state yields squeezing of the quantum field and, as a result,  entanglement between the two sub-systems is also produced~\cite{BUCHDADUMORU01}. 
It is of particular interest to see how the Husimi $Q$-function behaves once the Kerr medium is added. When $\chi/\lambda$ increases by a fraction value, we notice clearly the single blob become almost perfectly circular with radius $|\alpha|\approx 7.5$ rotates in the counterclockwise direction, see Figs. 8a, b, d and 8f. This behavior of the quantum field distribution is similar to that of the thermal state~\cite{SACHNEBU03} which means that once the Kerr medium is added, makes it clear that rethermalization of the binomial field is indeed taking place. Now, it is perfectly sensible to ask, what is the situation if the Kerr Medium parameter is increased by an integer value? In this case, when the state evolves further, a multi-component structure develops as shown in Figs 8c, e and 8g, respectively. In these figures, the Husimi $Q$-function demonstrates that the quantum states obtained corresponding to Schr\"{o}dinger cat states. Moreover, we see that the cat states have different number of components at different values of $\chi/\lambda$. Different mechanisms demonstrated that, for the case of a radiation field propagating in a nonlinear medium, Schr\"{o}dinger cat states are generated~\cite{FuSOLO001, MIMAWA00} with different number of components at different times in the evolution~\cite{DYAO97}. It was shown that the splitting of the Husimi function, which is the signature of the formation of Schr\"{o}dinger cat states, is related strongly to quantum entanglement~\cite{VAOR95, ORPAKA95, JEOR94, MIMAWA00, MIWAIM01}.\\
To discuss the evolution of the Husimi $Q$-function in the case of resonance and fixed value of the Kerr parameter, i.e., $\chi/\lambda=5.0$ (strong Kerr medium), we have setted various values for the Rabi angle $\lambda t$ in our computations, i.e., $\lambda t=0.0, \pi/6, \pi/4, \pi/3, \pi/2$ and $\pi$. The results are displayed in figure 9. A collision of six blobs occurs gradually when $\lambda t=\pi/6$, which implies the rethermalisation of the quantum field. At the time evolution of $\lambda t=\pi, \pi/2, \pi/3$ and $\pi/4$ the distribution of the Husimi $Q$-function splits into two, three and four blobs corresponding to Schr\"{o}dinger cat states corresponding to different number of components at different times in the evolution. The center of blobs lies on a circle with radius $|\alpha|$ centered at $X=Y=0.0$.\\
\newpage
\vspace{-.2cm}
\begin{figure}[tpbh]
\begin{center}
\includegraphics[width=0.3\linewidth]
{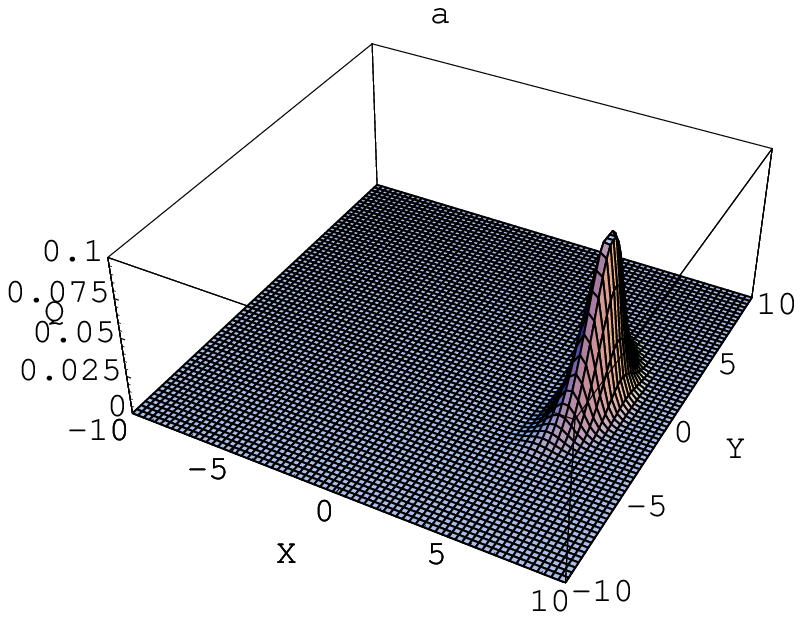}
\includegraphics[width=0.22\linewidth]
{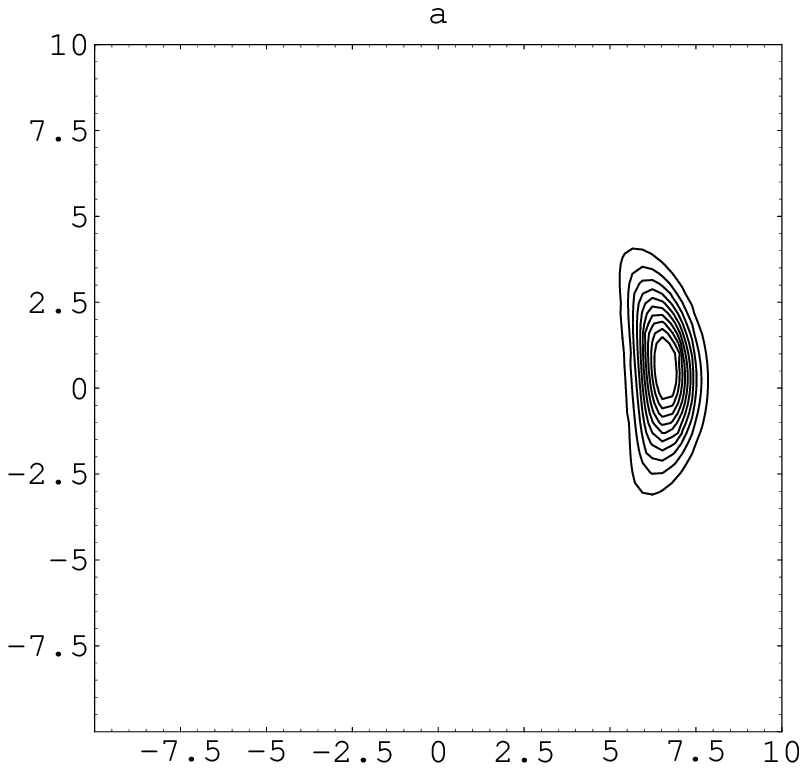}
\\
\includegraphics[width=0.3\linewidth]
{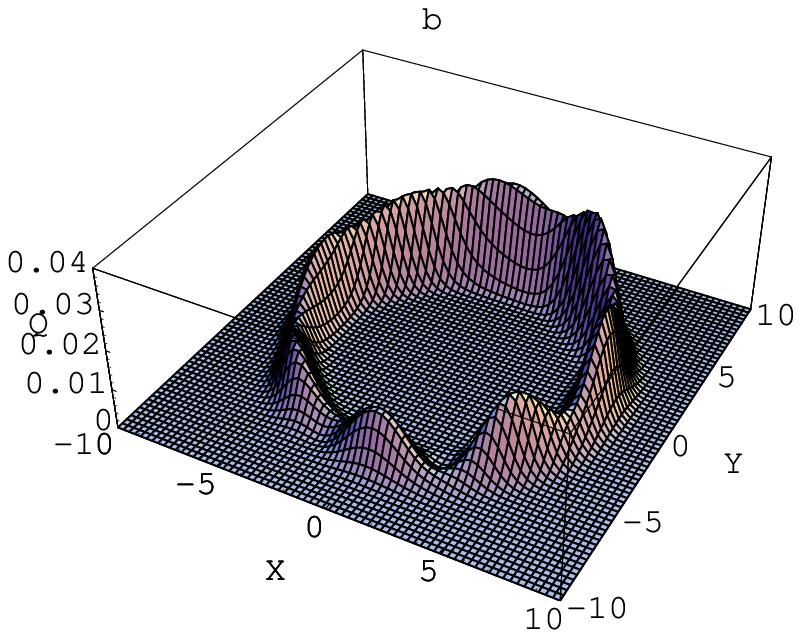}
\includegraphics[width=0.22\linewidth]
{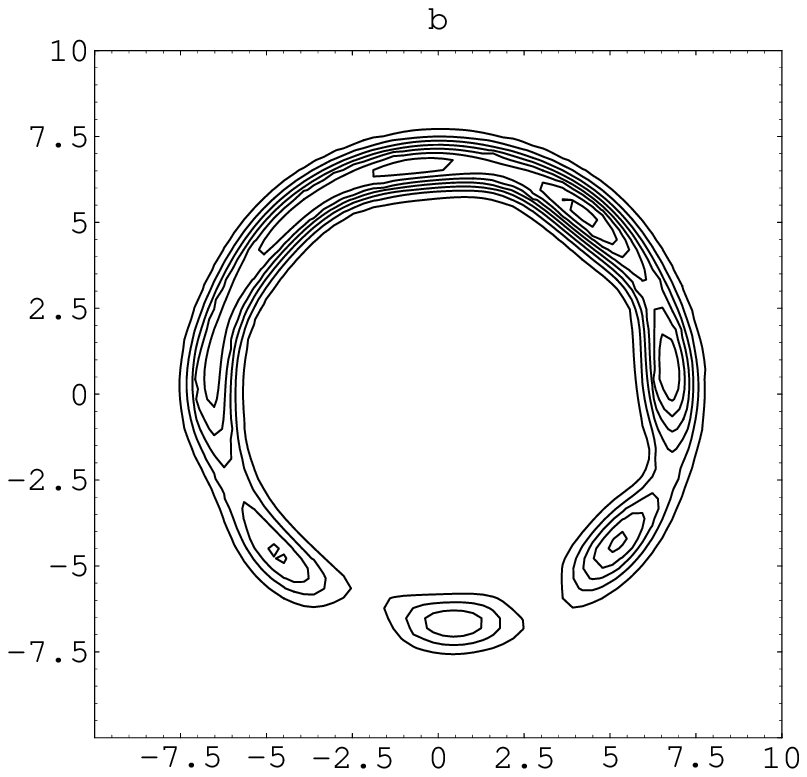}
\\
\includegraphics[width=0.3\linewidth]
{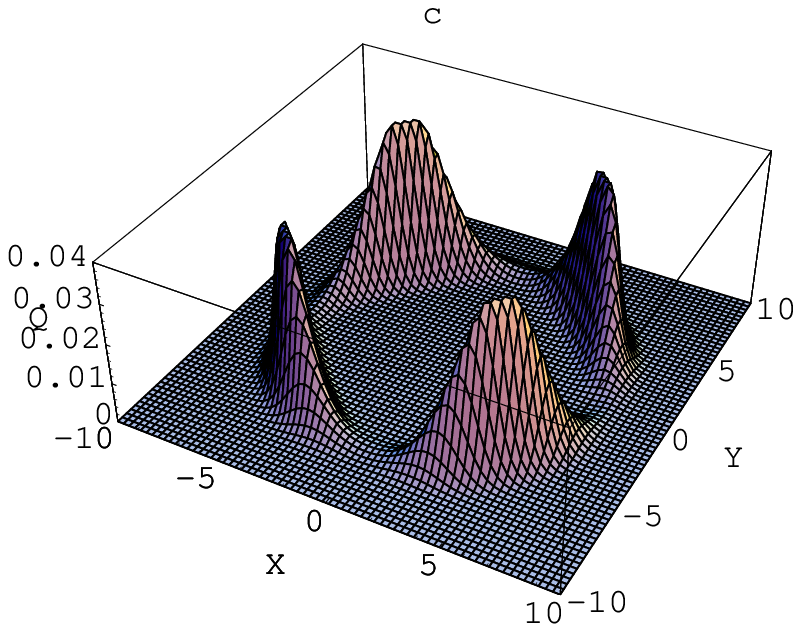}
\includegraphics[width=0.22\linewidth]
{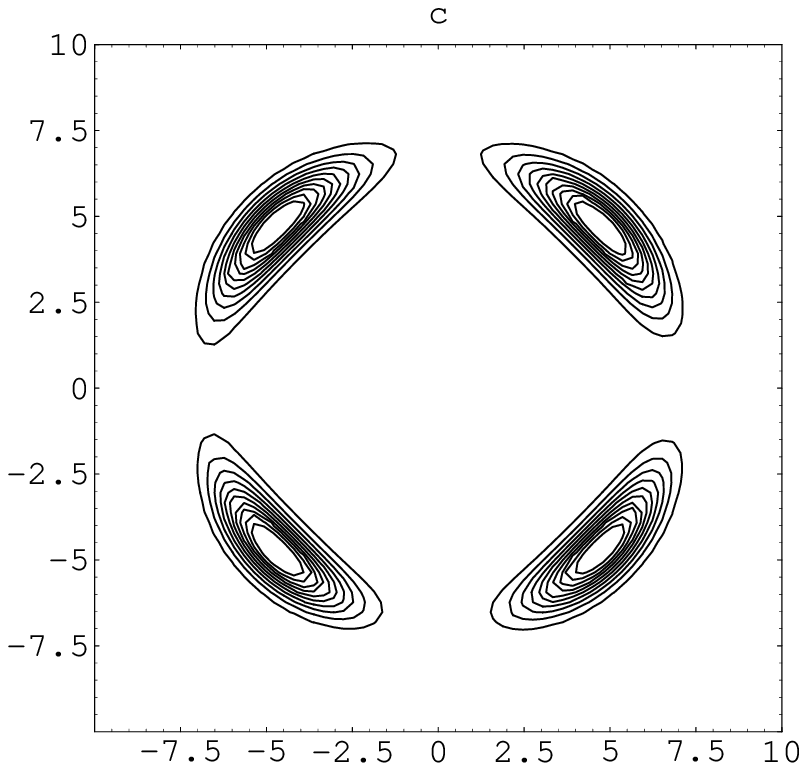}
\\
\includegraphics[width=0.3\linewidth]
{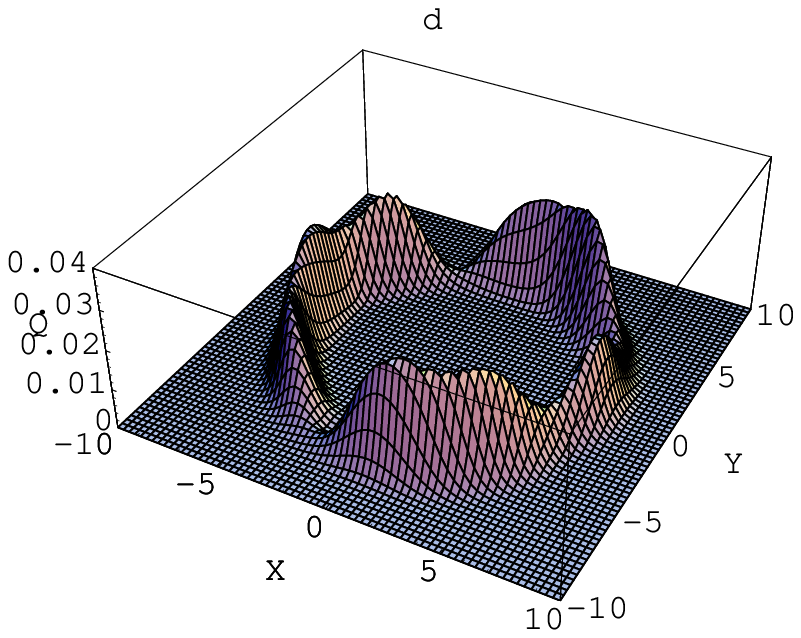}
\includegraphics[width=0.22\linewidth]
{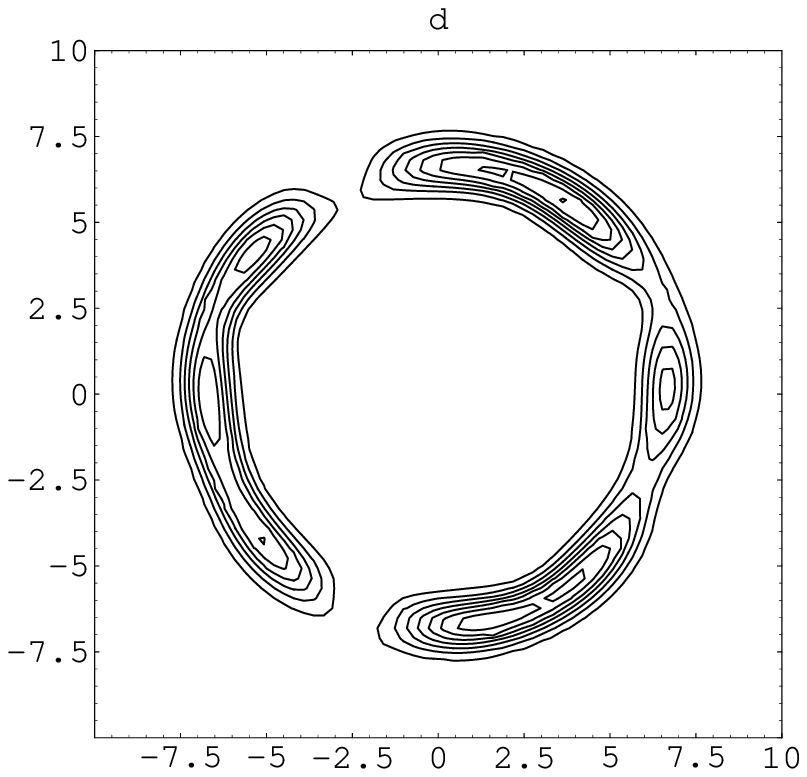}
\caption {Husimi Q-function with $|p|=0.9$, $M=50$,  $t=\pi/4$,
$\Delta/\lambda=0$ and (a)$\chi/\lambda=0.0$, (b)$\chi/\lambda=0.5$, (c) $\chi/\lambda=1.0$, (d) $\chi/\lambda=1.5$, (e) $\chi/\lambda=2.0$, (f) $\chi/\lambda=2.5$ and (g) $\chi/\lambda=5.0$ }
\end{center}
\end{figure}
\begin{figure}[tpbh]
\begin{center}
\includegraphics[width=0.3\linewidth]
{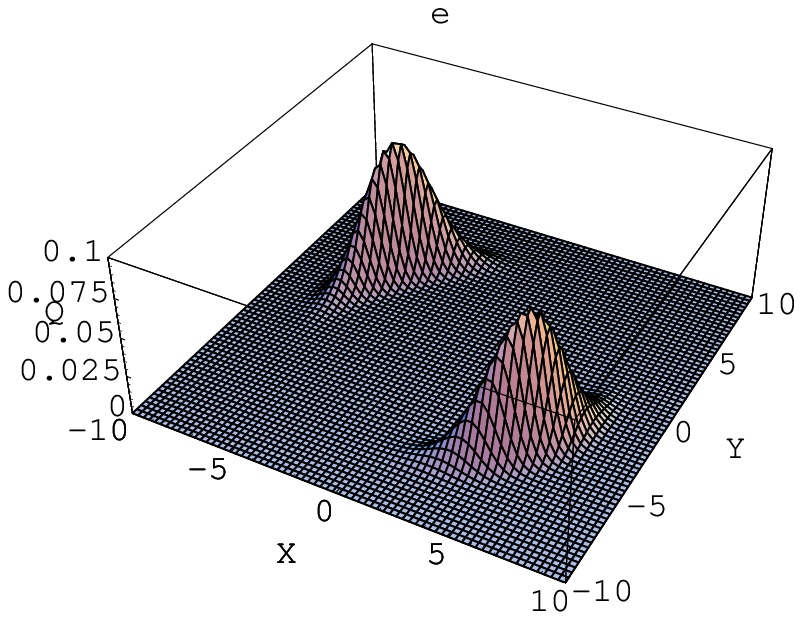}
\includegraphics[width=0.22\linewidth]
{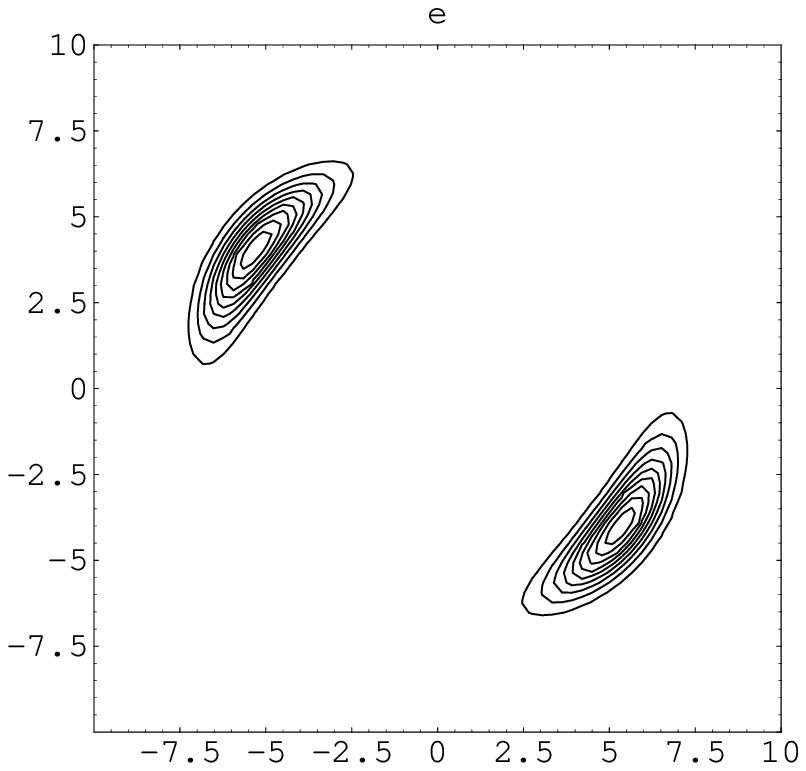}
\\
\includegraphics[width=0.3\linewidth]
{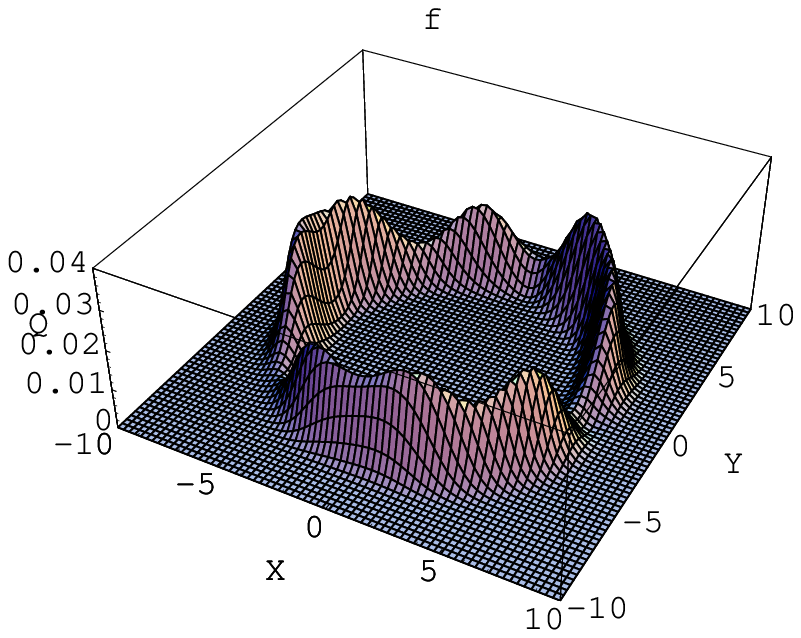}
\includegraphics[width=0.22\linewidth]
{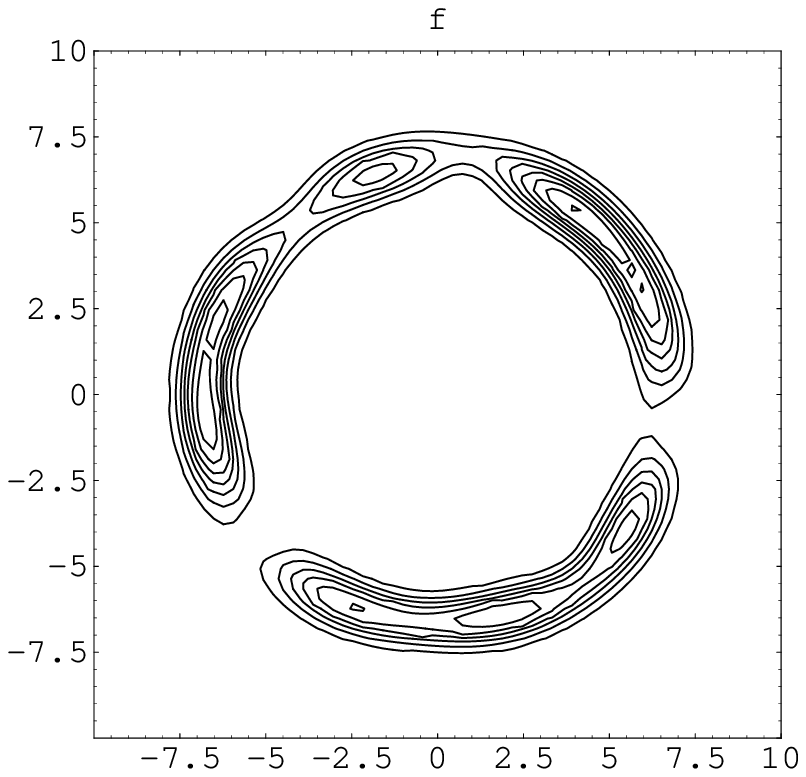}
\\
\includegraphics[width=0.3\linewidth]
{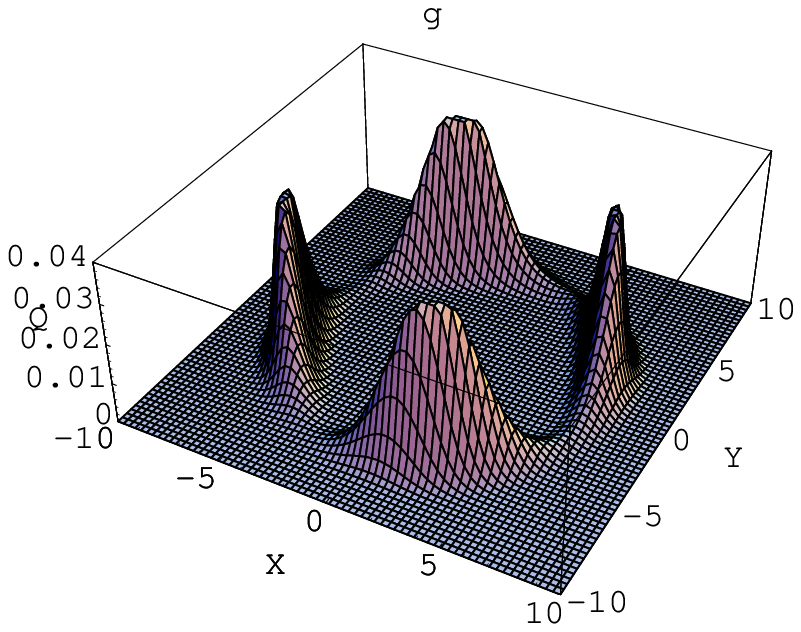}
\includegraphics[width=0.22\linewidth]
{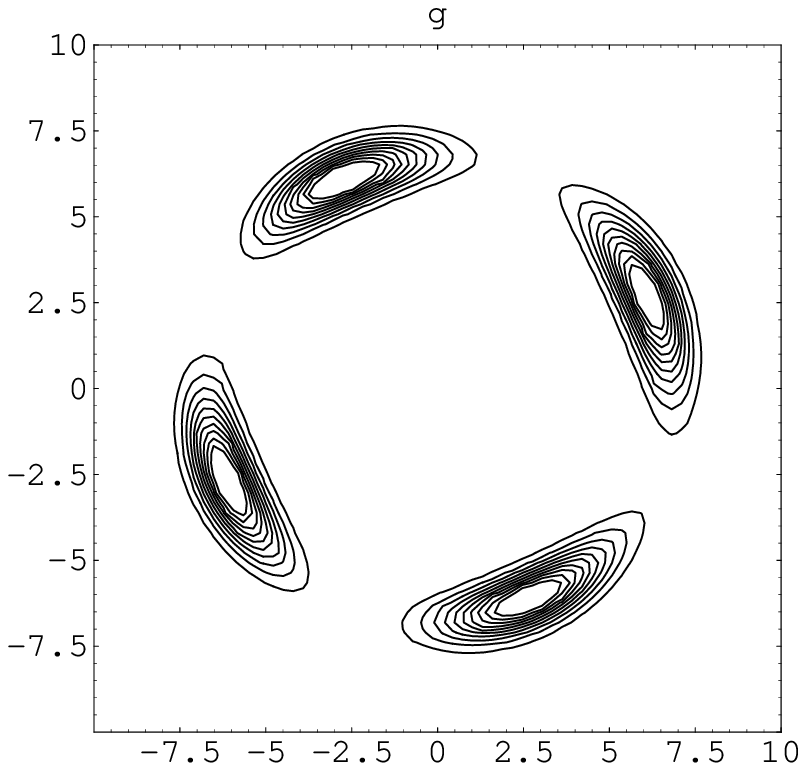}
\begin{center}
FIG. 8: continued
\end{center}
\end{center}
\end{figure}

\newpage
\begin{figure}[tpbh]
\begin{center}
\includegraphics[width=0.3\linewidth]
{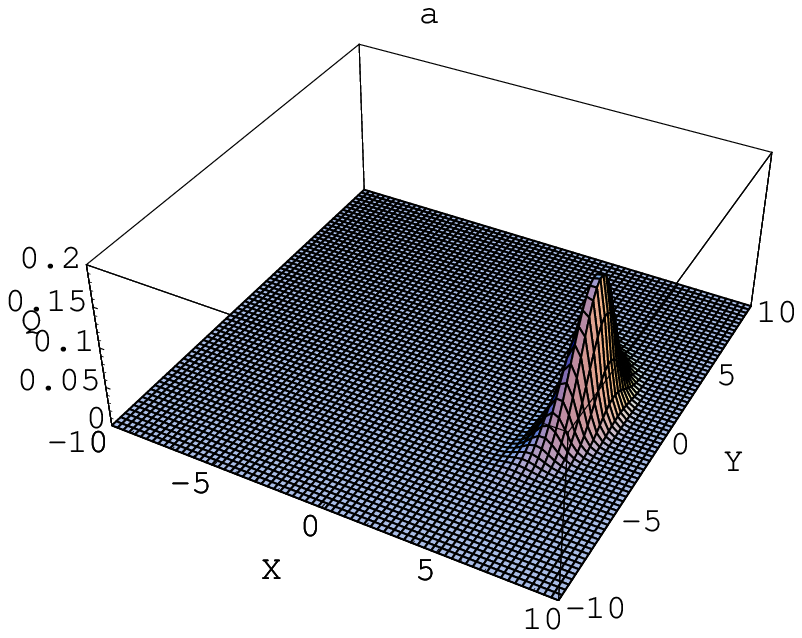}
\includegraphics[width=0.22\linewidth]
{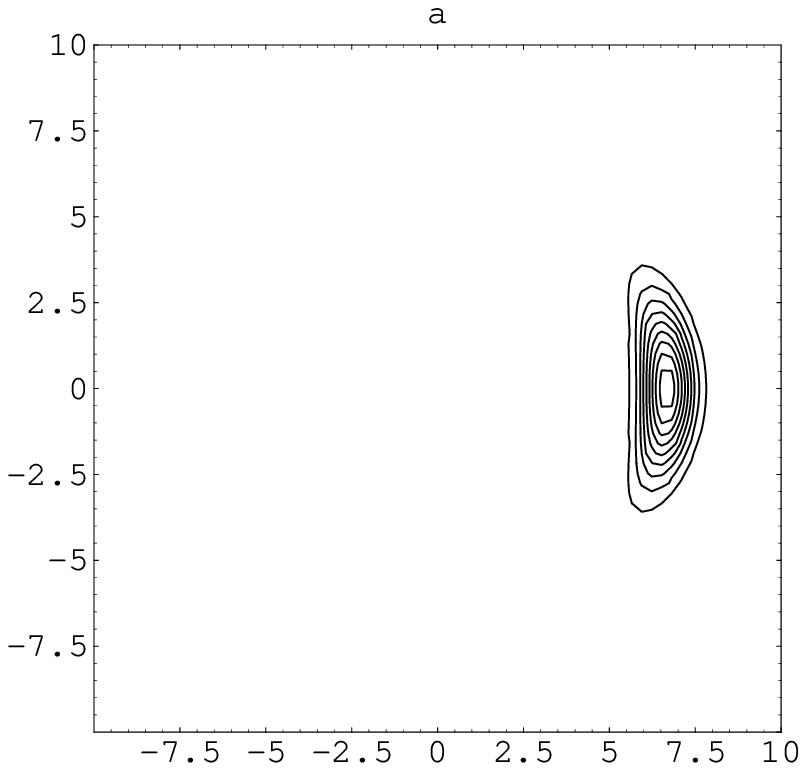}
\\
\includegraphics[width=0.3\linewidth]
{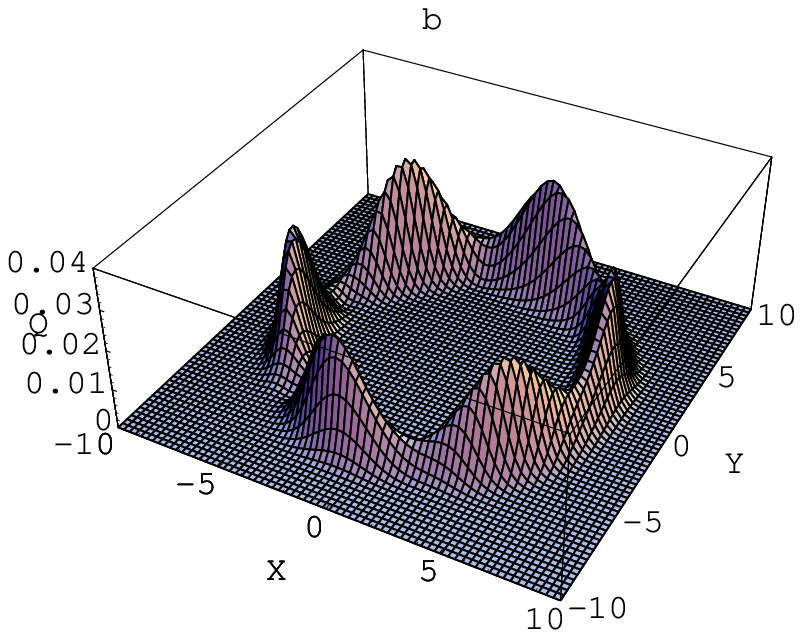}
\includegraphics[width=0.22\linewidth]
{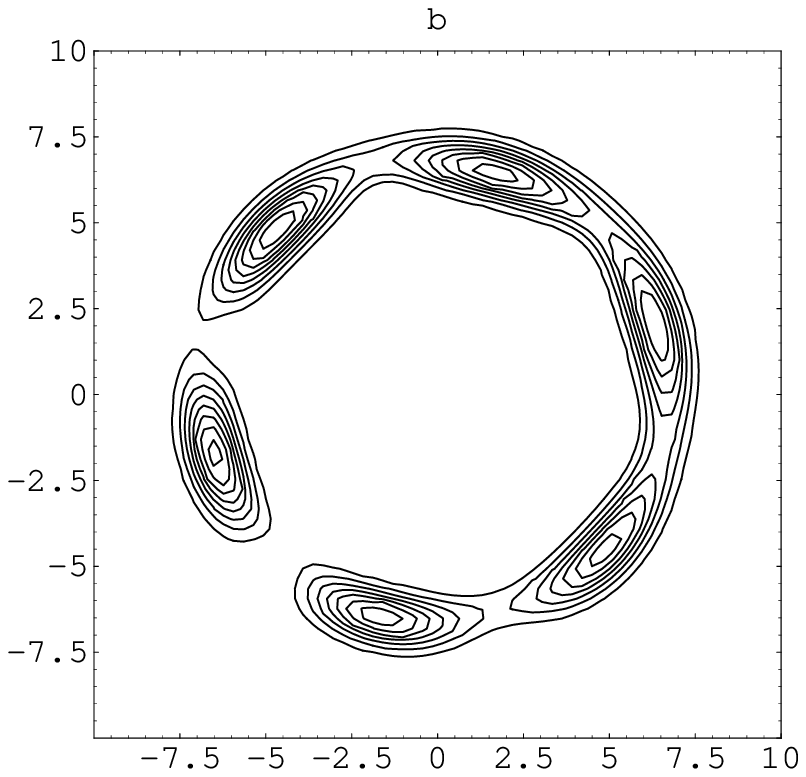}
\\
\includegraphics[width=0.3\linewidth]
{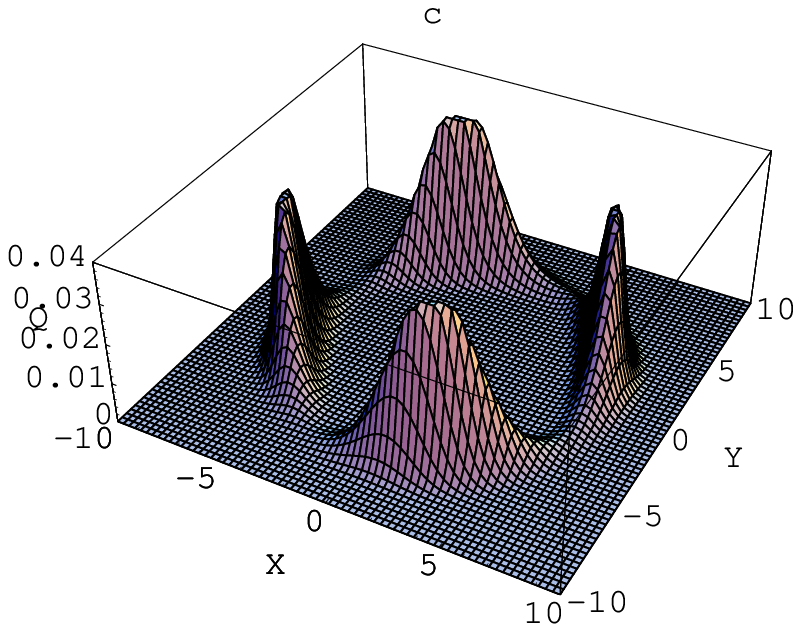}
\includegraphics[width=0.22\linewidth]
{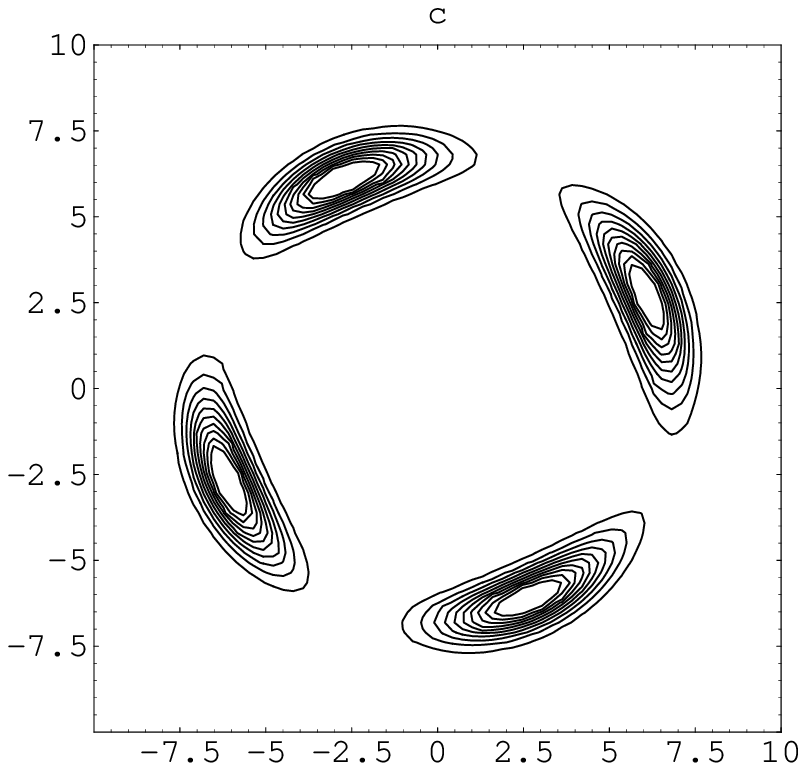}
\\
\includegraphics[width=0.3\linewidth]
{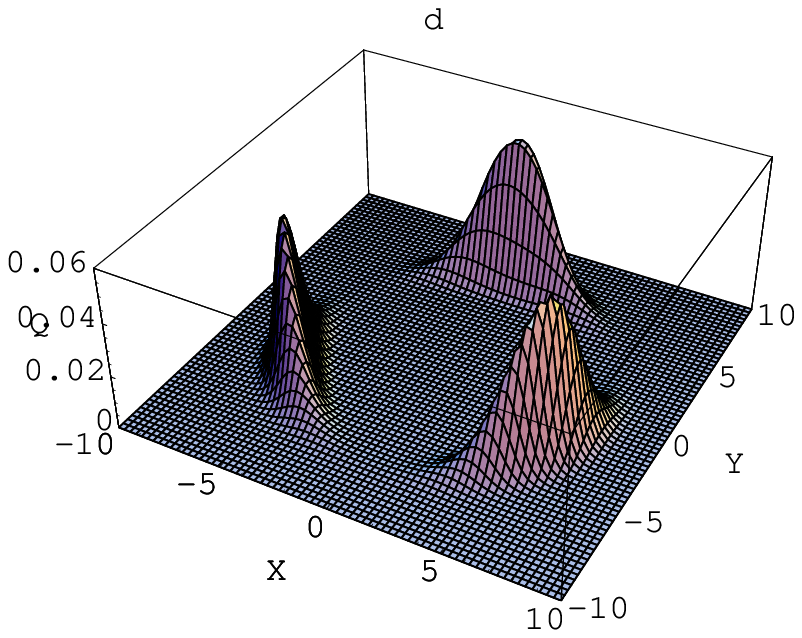}
\includegraphics[width=0.22\linewidth]
{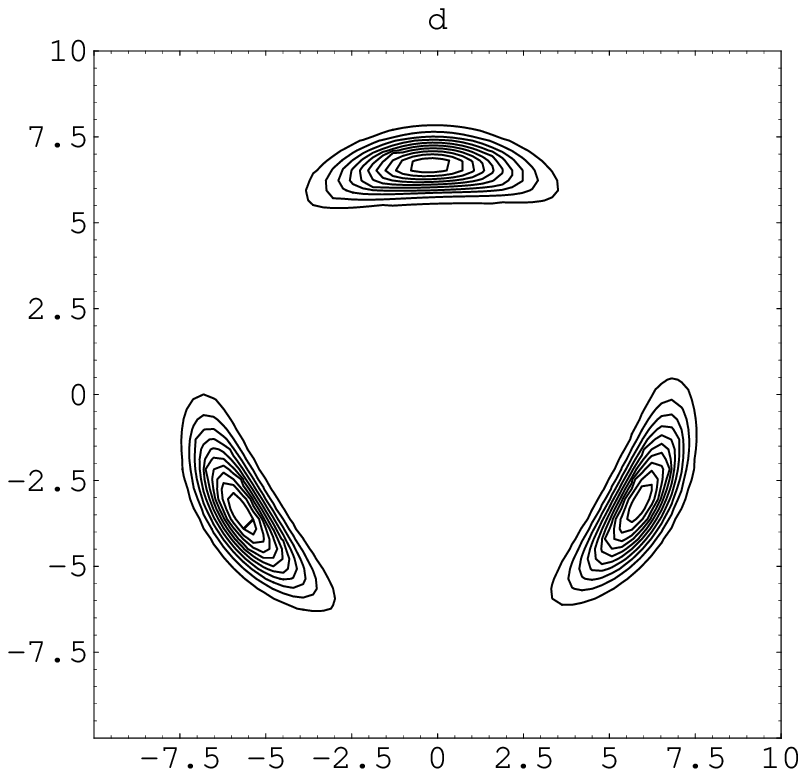}
\caption{Husimi Q-function with $|p|=0.9$, $M=50$,  $\chi=5.0$,
$\Delta/\lambda=0$ and (a) $t=0.0$, (b)$t=\pi/6$, (c) $t=\pi/4$, (d) $t=\pi/3$, (e) $t=\pi/2$ and (f) $t=\pi$ }
\end{center}
\end{figure}

\newpage
\begin{figure}[tpbh]
\begin{center}
\includegraphics[width=0.3\linewidth]
{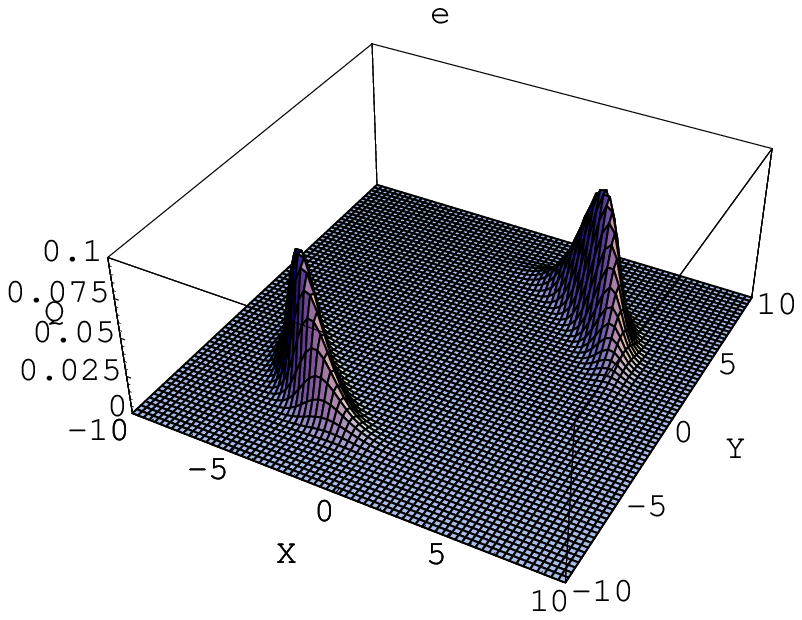}
\includegraphics[width=0.22\linewidth]
{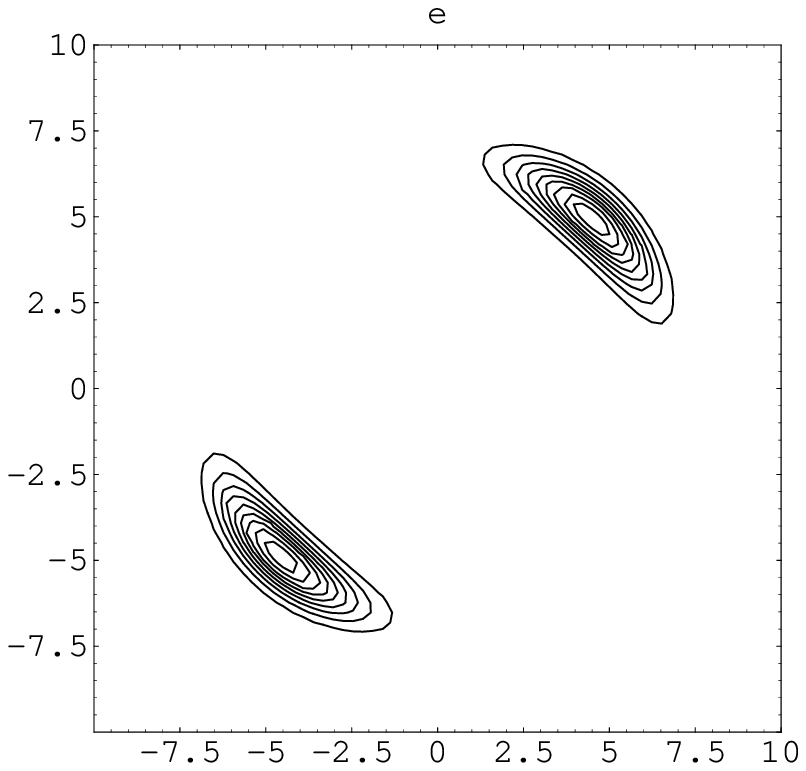}
\\
\includegraphics[width=0.3\linewidth]
{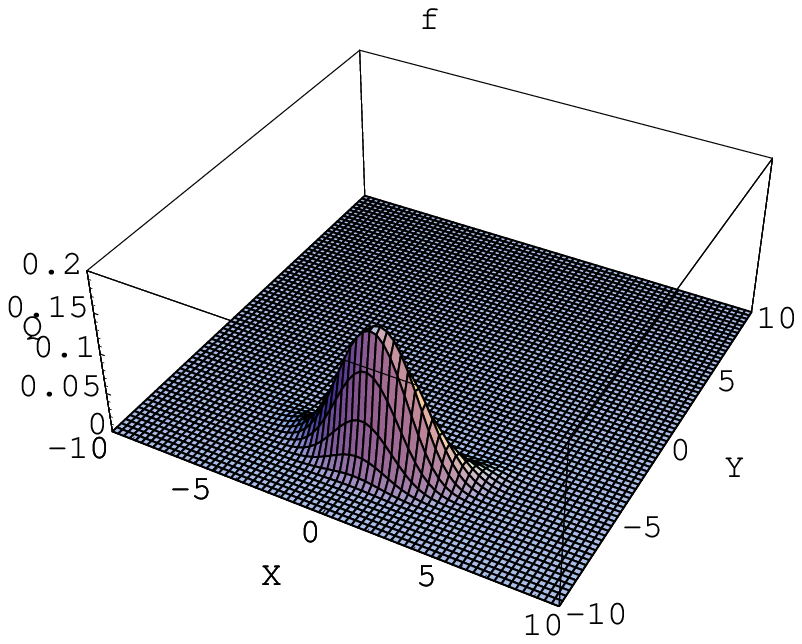}
\includegraphics[width=0.22\linewidth]
{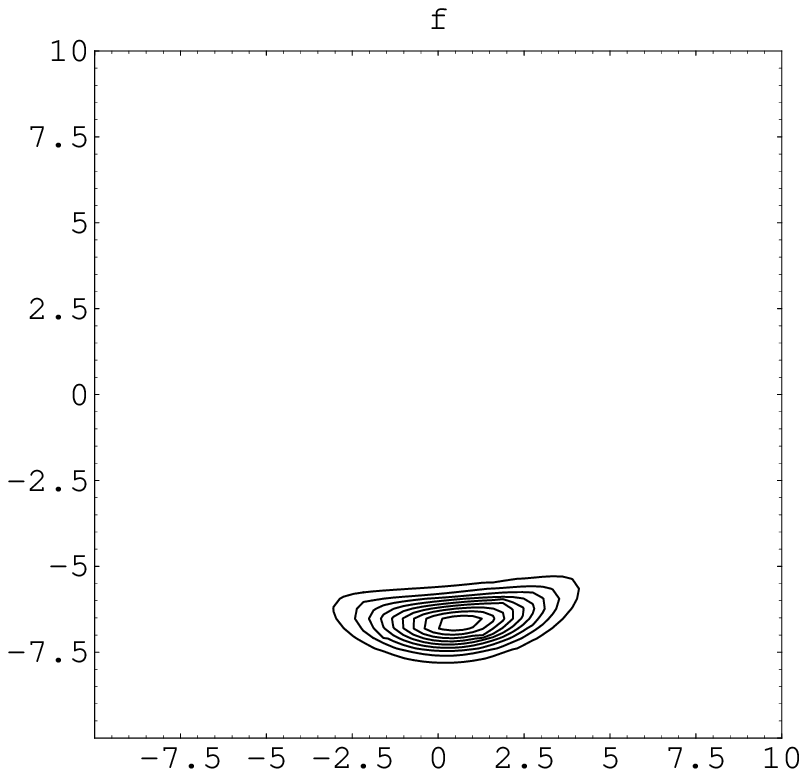}
\begin{center}
FIG. 9: continued
\end{center}
\end{center}
\end{figure}
%
%\newpage
%

\newpage
\section{Conclusion}
In Conclusion, we have shown that spin squeezing implies entanglement for quantum tripartite-state, where the subsystem includes the bipartite-state is identical. We have proved that spin squeezing parameters can be a convenient tool to give some insight into the mechanism of entanglement for the model under consideration. Moreover, a subsystem cavity field contains a nonlinear medium enhances noticeably the dynamics of entanglement specially when interacts with atomic subsystem off-resonantly. More clear insight into the relation between entanglement and the phase space distribution i.e., Husimi $Q$-function, is also illustrated. In this situation, the strong Kerr medium stimulates the creation of Schr\"{o}dinger cat states which is necessary for the generation of entanglement.
%\newpage
%===============================================================================

%
\end{document}